\documentclass[12pt,letterpaper]{article}

\usepackage{amsmath}
\usepackage{amsfonts}
\usepackage{amssymb}
\usepackage{graphicx}

\def\beq{\begin{equation}}
\def\eeq{\end{equation}}
\newcommand{\bea}{\begin{eqnarray}}
\newcommand{\eea}{\end{eqnarray}}

\def\Tdot#1{{{#1}^{\hbox{.}}}}

\def\mp{m_\phi}
\def\mc{m_\chi}
\def\mP{M_P}

\def\S{{\cal S}}
\def\Sm{{\cal S}_{\rm m}}
\def\zetam{\zeta_{\rm m}}
\def\zetar{\zeta_{\rm r}}
\def\zetarad{\zeta_{\rm r}}
\def\i{{\rm inf}}

\def\k{{\vec k}}

\def\co{{c_\theta}}
\def\si{{s_\theta}}
\def\x{{\xi}}

\begin{document}

\begin{flushright} {\footnotesize }  \end{flushright}
\vspace{5mm}
\vspace{0.5cm}
\begin{center}

\def\thefootnote{\fnsymbol{footnote}}

{\Large \bf Non-linear isocurvature perturbations}\\
\vspace{0.3cm} {\Large \bf and non-Gaussianities}
\\[0.5cm]
{\large David Langlois$^{\rm a}$, Filippo Vernizzi$^{\rm b}$
and David Wands$^{\rm c}$}
\\[0.5cm]

{\small \textit{$^{\rm a}$ APC (CNRS-Universit\'e Paris 7),\\ 10
rue Alice Domon et L\'eonie Duquet, 75205 Paris Cedex 13, France}}

\vspace{.2cm}

{\small \textit{$^{\rm b}$ Abdus Salam ICTP, Strada Costiera 11,
34014 Trieste, Italy}}

\vspace{.2cm}

{\small \textit{$^{\rm c}$ Institute of Cosmology and
Gravitation,\\ University of Portsmouth, Portsmouth, PO1 2EG,
United Kingdom}}

\end{center}

\vspace{.8cm}

\hrule \vspace{0.3cm}
{\small  \noindent \textbf{Abstract} \\[0.3cm]
\noindent We study non-linear primordial adiabatic and isocurvature
perturbations and their non-Gaussianity. After giving a general
formulation in the context of an extended $\delta N$ formalism, we
analyse in detail two illustrative examples. The first is a mixed
curvaton-inflaton scenario in which fluctuations of both the
inflaton and a curvaton (a light isocurvature field during
inflation) contribute to the primordial density perturbation. The
second example is that of double inflation involving two decoupled
massive scalar fields during inflation.
In the mixed curvaton-inflaton scenario we find that the bispectrum
of primordial isocurvature perturbations may be large and comparable
to the bispectrum of adiabatic curvature perturbations.

 \vspace{0.5cm} \hrule
\def\thefootnote{\arabic{footnote}}
\setcounter{footnote}{0} \vspace{0.5cm}

\section{Introduction}

With recent cosmic microwave background (CMB) anisotropy data due to
the WMAP satellite and the further improved data expected from the
Planck satellite, our knowledge of primordial cosmological
perturbations is becoming more and more precise. This influx of data
has stimulated the study of models whose predictions differ from the
simplest models of single field slow-roll inflation. To discriminate
between these models, a particularly important observable is the
amplitude (and the shape) of non-Gaussianity of the CMB
anisotropies. Another crucial property, potentially observable in
the CMB data, would be the presence of a primordial isocurvature (or
entropy) component as it would require a multi-field scenario for
the origin of the primordial fluctuations.

The purpose of this paper is to investigate the impact of
non-adiabatic fluctuations during inflation on the predicted
non-Gaussianity of primordial density perturbations, including
primordial isocurvature matter perturbations as well as adiabatic
modes which would contribute to the bispectrum and higher-order
correlations in the CMB anisotropies.
Isocurvature perturbations could have large departures from
Gaussianity while remaining sub-dominant in the linear perturbation
spectrum \cite{Linde:1996gt,Boubekeur:2005fj}.

To study primordial non-Gaussianity, one needs to study
relativistic cosmological perturbations beyond linear order and
there has been considerable progress in this field in recent
years. On scales larger than the Hubble radius, the non-linear
evolution of perturbations generated during  inflation is
compactly described in terms of the perturbed expansion from an
initial hypersurface (usually taken at Hubble crossing during
inflation) up to a final uniform-density hypersurface (usually
during the radiation-dominated era) -- the so-called $\delta
N$-formalism \cite{deltaN}. This is particularly useful for
evaluating the primordial non-Gaussianity generated on large
scales \cite{Lyth:2005fi}.

As we show in this paper, one can easily extend the $\delta
N$-formalism to describe the non-Gaus\-sia\-ni\-ties of the
non-linearly evolved primordial perturbations including
isocurvature fluctuations. In order to illustrate our general, but
formal, result, we study two emblematic examples of multi-field
scenarios, which can generate isocurvature fluctuations in
addition to the usual adiabatic fluctuations. The first example is
the curvaton scenario \cite{Linde:1996gt,curvaton}.
Previous works, e.g.~\cite{Lyth:2002my,Gordon:2002gv,beltran}, have
investigated non-Gaussianity and isocurvature perturbations in
this scenario, but in our case,
we do not assume that the contribution of inflaton fluctuations to the CMB anisotropies is negligible.
In this
so-called mixed inflaton-curvaton setup
\cite{Langlois:2004nn,Ferrer:2004nv}, the isocurvature mode is not
necessarily constrained by the data to be zero, in contrast with
the conclusion of \cite{beltran}.
Our second example is a model of
double inflation \cite{Polarski:1992dq} with two uncoupled massive
scalar fields that drive in turn inflation. In contrast with the
previous example, the final isocurvature perturbation depends on
both scalar field fluctuations during inflation, but it can still
be determined analytically at second order.

The adiabatic and isocurvature perturbations we refer to above
correspond to the primordial adiabatic and isocurvature
perturbations defined during the standard radiation era, i.e.,
after inflation and {\it after} the curvaton decay, if any. These
perturbations can be related, but are not equivalent, to the {\it
instantaneous} adiabatic and isocurvature (or entropy) field
perturbations which can be defined during inflation by decomposing
the perturbations along the directions, respectively, parallel and
orthogonal to the inflationary trajectory in field space (see
\cite{Gordon:2000hv,GrootNibbelink:2001qt} in the linear case and
\cite{NLdec,Langlois:2006vv} in the non-linear case). The
instantaneous isocurvature perturbation
during inflation
is not necessarily converted
into an isocurvature perturbation after inflation. However, even
if the isocurvature perturbation during inflation does not
survive, it can have a strong impact on the resulting primordial
adiabatic perturbation and its non-Gaussianity, as illustrated,
for instance,
recently in the context
of multi-field Dirac-Born-Infeld inflation \cite{DBI}.

The outline of the paper is the following. In the next section, we
introduce the non-linear definitions of the primordial adiabatic and
isocurvature perturbations and show how they are related to the
primordial scalar field fluctuations in a very general multi-field
inflation framework. The following section is  devoted to the study
of the mixed curvaton-inflaton scenario. We then consider, in the
fourth section, the case of double inflation with two decoupled
massive scalar fields. We discuss our results in the final section.
In Appendix A we give some details of the calculations of section IV,
while in Appendix B we review the
decomposition into the adiabatic and entropy components of the field
perturbations and  their equations of motion at second order, and we
compute their
3-point correlation functions. Finally, in the last appendix we
give general expressions obtained using the $\delta N$-formalism for the
primordial power spectra and bispectra at leading order.

While this paper was being written up, similar results have been
obtained by Kawasaki {\em et al.}~who use the $\delta N$-approach to
calculate primordial non-Gaussianity of isocurvature perturbations
and in particular axion isocurvature perturbations
\cite{Kawasaki:2008sn}
and baryon isocurvature perturbations \cite{Kawasaki:2008jy}.

\section{Adiabatic and isocurvature perturbations}

A powerful technique to compute the non-linear primordial
perturbations on large scales is the $\delta N$-formalism
\cite{deltaN,Lyth:2005fi}. The idea is to use solutions to the
homogeneous FRW cosmology in order to calculate the integrated
expansion on large scales from some initial state to a final state
of fixed energy density.

The $\delta N$-formalism is  closely related to the notion of non-linear curvature perturbation
on uniform density hypersurfaces, which can be defined in a geometrical and covariant way as shown in
\cite{Langlois:2005ii,Langlois:2005qp}.
Indeed,
in the case of a perfect fluid characterized by the energy density $\rho$, the pressure $P$ and the four-velocity $u^a$, the conservation law for the energy-momentum tensor,
 \beq
\nabla_a T^{a}_{\ b}=0, \qquad T_{ab}=\left(\rho+P\right) u_a u_b
+P g_{ab},
 \eeq
implies that the covector
\beq
 \zeta_a\equiv
\nabla_a\alpha-\frac{\dot\alpha}{\dot\rho}\nabla_a\rho
\label{zeta_a}
\eeq
satisfies the relation
\beq
\label{dot_zeta}
\dot\zeta_a\equiv {\cal L}_u\zeta_a=
-\frac{\Theta}{3(\rho+p)}\left( \nabla_a p -
\frac{\dot p}{\dot \rho} \nabla_a\rho\right) \;,
\eeq
where we have defined
\beq
 \Theta=\nabla_a u^a, \quad \alpha=\frac{1}{3}\int d\tau \,
\Theta \;,
\eeq
and where a dot denotes a Lie derivative along $u^a$, which is equivalent to an ordinary
derivative for {\it scalar} quantities (e.g. $\dot\rho\equiv u^a\nabla_a\rho$). This result is valid for any spacetime geometry and does not depend on Einstein's equations. In the cosmological context, $\alpha$ can be interpreted as a non-linear generalization, according to an observer following the fluid, of the number of e-folds of the scale factor.

The covector $\zeta_a$ can be defined for the global cosmological fluid or for any of the
individual cosmological fluids (the case of interacting fluids is discussed in \cite{Langlois:2006iq}). Using
the non-linear conservation equation
\beq
\dot\rho=-3\dot\alpha(\rho+P)\;,
\eeq
which follows from $u^b\nabla_a T^a_{\ b}=0$,
one can re-express  $\zeta_a$ in the form
\beq
\label{zeta_a2}
\zeta_a=\nabla_a\alpha+\frac{\nabla_a\rho}{3(\rho+P)} \;.
\eeq
If $w\equiv P/\rho$ is constant,
the above covector  is a total gradient and can be written as
\beq
\zeta_a=\nabla_a\left[\alpha+\frac{1}{3(1+w)}\ln \rho\right]\, .
\label{zeta_a_w}
\eeq

On scales larger than the Hubble radius, the above definitions are
equivalent to the non-linear curvature perturbation on uniform
density hypersurfaces as defined in
\cite{Lyth:2004gb,Sasaki:2006kq},
 \beq \zeta = \delta N -
 \int_{\bar \rho}^{\rho} H \frac{d \tilde \rho}{\dot{ \tilde \rho}}
 =
 \delta N + \frac13\int_{\bar \rho}^{\rho} \frac{d
 \tilde{\rho}}{(1+w)\tilde{ \rho}}\;, \label{zeta}
 \eeq
where $N=\alpha$ and  $H=\dot\alpha=\dot a /a$ is the Hubble rate
of the Friedmann metric $ds^2 = -dt^2 +a^2(t) d\vec x^2$. The
above equation is simply the  integrated version of
(\ref{zeta_a}), or of (\ref{zeta_a2}).

In the following, we will be mainly interested in   non-linear
isocurvature, or entropy, perturbations. For simplicity, we will
consider only  cold dark matter (CDM) isocurvature perturbations and
assume that the Universe, in the standard eras, is filled with  only
two fluids: the radiation fluid and the CDM fluid. Our analysis can
be easily extended to other types of isocurvature perturbations.

It will be useful to distinguish  the non-linear curvature
perturbation $\zeta$ of the total fluid, which describes the
primordial {\it adiabatic} perturbation, from the non-linear
perturbations $\zetar$ and $\zetam$ describing respectively the
radiation  fluid ($w_r=1/3$) and the cold dark matter (CDM) fluid
($w_m=0$), which are given,
according to our definitions (\ref{zeta_a_w}) or (\ref{zeta}),
 by
 \beq
 \label{defzetar}
  \zetar = \delta N +
\frac14 \ln\left( \frac{\rho_{\rm r}}{\bar
    \rho_{\rm r}} \right)\; ,
 \eeq
 \beq
\zetam = \delta N + \frac13 \ln\left( \frac{\rho_{\rm m}}{\bar
    \rho_{\rm m}} \right)\; , \label{zetadust}
 \eeq
where a bar denotes a homogeneous quantity.

In the radiation dominated era, the adiabatic perturbation
coincides with
$\zetar$, whereas the CDM isocurvature perturbation is
characterized
by the non-linear perturbation
 \beq
 \label{defSm}
  \Sm = 3(\zetam - \zetar)
   = \ln\left( \frac{\rho_{\rm m}}{\bar \rho_{\rm
m}} \right) - \frac34 \ln\left( \frac{\rho_{\rm r}}{\bar{
\rho}_{\rm r}} \right)\; ,
 \eeq
which can be expanded in terms of the density contrasts
$\delta_{\rm r} = \delta \rho_{\rm r} / \bar{ \rho}_{\rm r} $ and
$\delta_{\rm m} = \delta \rho_{\rm m} / \bar{\rho}_{\rm m} $,
 \beq
  \Sm =
 \delta_{\rm m}  - \frac34 \delta_{\rm r} - \frac12 \delta^2_{\rm m}
+ \frac38 \delta^2_{\rm r} + \ldots
 \eeq
Note that these expressions are independent of the hypersurface on
which the density perturbations are defined.

Since our goal is to relate the perturbations in the radiation era
to the perturbations produced during an inflationary era, it is
important to generalize Eq.~(\ref{zeta}) for scalar fields. In
this case a convenient description is in terms of the (relative)
comoving curvature perturbation,\footnote{Note that the convention
adopted here is that
${\cal R}_A$ has the same sign as $\zeta_A$,
such that, in the single field case,  ${\cal R}=\zeta$ on large scales.}
\beq
 {\cal R}_A = \delta N - \int_{\bar \varphi_A}^{\varphi_A} H
\frac{d \tilde{\varphi}_A}{\dot {\tilde{\varphi}}_A}\;,
\label{zeta_field}
 \eeq
which is the curvature perturbation on constant $\varphi_A$ hypersurface.
In slow-roll inflation, the initial state of
the system -- when the cosmological perturbations are produced --
is defined only by the scalar field values, $\varphi_{A*}$, on an
initial spatially-flat hypersurface,
where with a star we denote that we evaluate the quantity at
Hubble crossing $k=aH$. One can then calculate the
number of e-folds, or integrated expansion, $N^{(\varphi_A)}$,
from this initial state to a ``final'' hypersurface characterized
by the ``final'' scalar field amplitudes $\varphi_A$. By choosing
the final hypersurface to be of uniform $A$-field, one can write
${\cal R}_A$ as a perturbative expansion in terms of the initial
field fluctuations $\delta \varphi_{A*}$, whose correlation
properties must be known. Equation (\ref{zeta_field}) thus becomes
 \beq {\cal R}_A  = \delta N^{(\varphi_A)}
 = N^{(\varphi_A)}_{,A}   \delta \varphi_{A*} + \frac{1}{2}
 N^{(\varphi_A)}_{,AB} \delta \varphi_{A*} \delta
 \varphi_{B*} + \ldots\;,
  \label{zetaI_N}
 \eeq
where $N^{(\varphi_A)}_{,A} = \partial N^{(\varphi_A)}/\partial
\varphi_{A*}$, etc. This is a particular application of the
$\delta N$-formalism that generalizes the usual expansion of $N$
defined on a final {\em total} uniform density hypersurface. Note
that when there are several scalar fields, ${\cal R}_A$ can be
different from the relative curvature perturbation on uniform
density hypersurfaces $\zeta_A$. Indeed, the uniform density and
uniform field hypersurfaces do not always coincide even on large
scales \cite{Polarski:1994rz}.

However, the {\em total} comoving and uniform density
hypersurfaces coincide on large scales at second \cite{RC} and
non-linear order \cite{Lyth:2004gb,Langlois:2006vv} and $\zeta$ is
generally used to describe the adiabatic perturbation also for
scalar fields. The curvature perturbation on uniform density
hypersurfaces
$\zeta$ will be given now as the standard perturbative expansion
of $N$ defined on a final uniform density hypersurface. Thus one
can rewrite $\zeta$ in terms of the expansion \cite{Lyth:2005fi}
 \beq
\zeta  = \delta N =  N_{,A}  \delta \varphi_{A*} + \frac{1}{2}
N_{,AB}  \delta \varphi_{A*} \delta \varphi_{B*} +\ldots \;.
\label{zeta_N}
 \eeq
Similarly, the non-linear isocurvature perturbation (\ref{defSm})
can be given in terms of the difference in the non-linear
expansion,
$\Sm=3\delta(\Delta N)$,
where $\Delta N\equiv N^{\rm (m)} -  N^{\rm (r)}$,
between final hypersurfaces of uniform matter density and uniform
radiation density
 \beq
\Sm  = 3 \left( \delta N^{\rm (m)} - \delta N^{\rm (r)} \right)
 = 3 \Delta N_{,A} \delta \varphi_{A*} + \frac{3}{2} \Delta N_{,AB}
  \delta \varphi_{A*} \delta \varphi_{B*} +\ldots \;.
\label{Sm_N}
 \eeq

In the following sections we apply these definitions to two
examples: the curvaton model and double inflation. Although the
previous expressions hold non-linearly, we will concentrate on a
second order expansion, which is expected to give the leading
order terms for the 3-point correlation properties.
We will assume that the initial field perturbations (on scales close
to the horizon scale during inflation) are independent, Gaussian
random fields. Thus any non-Gaussianity of the curvature
perturbations will arise from the non-linear terms in
Eqs.~(\ref{zeta_N}) and~(\ref{Sm_N}). This is a good approximation
for weakly-coupled scalar fields (with canonical kinetic terms)
during slow-roll inflation \cite{Seery:2005gb} but may break-down
for scalar fields with non-standard kinetic terms.

\section{Mixed inflaton and curvaton perturbations}
\def\curv{\chi}

As a first application of the general formalism presented in the
previous section, we  consider a curvaton scenario \cite{curvaton},
or more precisely a mixed inflaton and curvaton scenario
\cite{Langlois:2004nn,Ferrer:2004nv} as we will take into account
both the perturbations  generated by the inflaton
field driving inflation
and the curvaton. The curvaton is a weakly coupled scalar field,
$\curv$,
which is light relative to the Hubble
rate during inflation, and hence acquires an almost
scale-invariant spectrum and effectively Gaussian distribution of
perturbations, $\delta\curv$, during inflation. After inflation
the Hubble rate drops and eventually the curvaton becomes
non-relativistic so that its energy density grows relative to
radiation, until it contributes a significant fraction of the
total energy density, $\Omega_\curv\equiv\bar\rho_\curv/\bar\rho$,
before it decays. Hence the initial curvaton field perturbations
on large scales can give rise to a primordial density perturbation
after it decays.

The non-relativistic curvaton (mass $m\gg H$), before it decays,
can be described by a pressureless, non-interacting fluid with
energy density
 \begin{equation}
\rho_\curv = m^2 \curv^2 \;,
 \end{equation}
where $\curv$ is the rms amplitude of the curvaton field, which
oscillates on a timescale $m^{-1}$ much less than the Hubble time
$H^{-1}$. Making use of Eq.~(\ref{zetadust}) for the oscillating
curvaton to rewrite its local density in terms of its homogeneous
value and the inhomogeneous expansion perturbation,
$\delta N$, we have
 \begin{equation}
  \rho_\curv = \bar\rho_\curv e^{3(\zeta_\curv-\delta N)} \label{rho_curvaton}\;.
 \end{equation}
In the post-inflation era where the curvaton is still subdominant, the spatially flat
hypersurfaces are characterized by $\delta N=\zeta_\i$, where $\zeta_\i$ corresponds to the
adiabatic perturbation generated by the inflaton fluctuations. On such a hypersurface, the
curvaton energy density can be written as
 \begin{equation}
 \bar\rho_\curv e^{3(\zeta_\curv-\zeta_\i)}=\bar\rho_\curv e^{\S_\curv}
 =
 m^2 \left( \bar\curv+\delta\curv \right)^2 \,. \label{rhorhobarcurv}
 \end{equation}
where
$\S_\curv\equiv 3(\zeta_\curv-\zeta_\i)$
is the entropy perturbation
of the curvaton.

As long as the curvaton is subdominant and weakly-coupled, so that
we may neglect self-interactions, the evolution equation for
$\curv$ is linear, and the field perturbation $\delta \curv$ obeys
the same evolution equation on super-Hubble scales as the
background expectation value $\bar \curv$. In this case it is well
known that the ratio $\delta \curv/\bar \curv$ remains unchanged
as long as the curvaton is subdominant \cite{Polarski:1992dq}.
This result holds also at second order in the perturbation $\delta
\curv$, as shown in \cite{Langlois:2006vv} and in
Appendix~B,
where we have written the evolution equation of an entropy field
perturbation. Thus, expanding Eq.~(\ref{rhorhobarcurv}) at second
order we obtain
 \beq
\S_\curv= 2\frac{\delta\curv_*}{\bar\curv_*}
-\left(\frac{\delta\curv_*}{\bar\curv_*} \right)^2 \,.
 \label{S_dchi}
 \end{equation}
Note that we will assume that the initial
curvaton field perturbations, $\delta\curv_*$, are strictly
Gaussian, as would be expected for a weakly coupled field.

The precise density perturbation produced after the curvaton
decays can be calculated numerically
\cite{Malik:2002jb,Malik:2006pm,Sasaki:2006kq}, but it can also be
estimated analytically using the sudden-decay approximation
\cite{Lyth:2002my}, which assumes that the curvaton decays
suddenly on a spatial hypersurface of uniform total energy
density.
Any initial inflaton perturbation gives rise to a perturbation in
the radiation energy density {\it before} the curvaton decay, which we denote
by $\rho_R$.
Similarly to Eqs.~(\ref{defzetar}) and
(\ref{zetadust}) we can write
\beq
\rho_R = \bar\rho_R e^{4(\zeta_\i-\delta N)} \, , \qquad \rho_\chi
 = \bar\rho_\chi
e^{3(\zeta_\chi-\delta N)} \,.
 \eeq
On the decay hypersurface characterized by $\rho_\chi+\rho_R=\bar
\rho_{\rm r}$ and thus $\delta N = \zetar$ where $\zetar$ is the total curvature perturbation {\it after} the
decay, we find
\cite{Sasaki:2006kq}
 \begin{equation}
 \label{decaysurface}
 \Omega_{\curv,{\rm decay}} e^{3(\zeta_\curv-\zetar)} +
 (1-\Omega_{\curv,{\rm decay}})
 e^{4(\zeta_\i-\zetar)} = 1 \;,
\end{equation}
where $\Omega_{\curv,{\rm decay}}\equiv \bar\rho_\chi/(\bar\rho_\chi+\bar\rho_{R})$.

Expanding Eq.~(\ref{decaysurface}) at first order we obtain
 \begin{equation}
 \zetar = r \zeta_\curv +(1-r)\zeta_\i\,,
 \end{equation}
where $r \equiv 3\Omega_{\curv,{\rm decay}}/(4-\Omega)_{\curv,{\rm
decay}}$.
Up to second-order we obtain
 \begin{equation}
 \zetar = r \zeta_\curv +(1-r) \zeta_\i+ \frac{r(1-r)(3+r)}{2}
 \left(\zeta_\curv-\zeta_\i\right)^2 = \zeta_\i+\frac{r}{3} {\cal S}_\curv+
\frac{r(1-r)(3+r)}{18}{\cal S}_\curv^2\,.
 \label{zeta_rad}
 \end{equation}
The entropy perturbation (\ref{S_dchi}) contains a linear part $S_G$
which is Gaussian and a second order part which is quadratic in
$S_G$:
 \beq
 {\cal S}_\curv=S_G-\frac14 S_G^2\, ,
  \quad {\rm where}\quad
 S_G \equiv 2\frac{\delta\curv_*}{\bar\curv_*}\, .
 \eeq
Substituting in (\ref{zeta_rad}) we then have
 \beq
 \zetarad =  \zeta_\i+\frac{r}{3} S_G+ \frac{r}{18}\left(\frac32-2r-r^2\right)S_G^2 \,.
 \label{zeta_rad3}
 \end{equation}
Keeping only the linear part of the above relation, one finds that
the power spectrum for the primordial adiabatic perturbation
$\zetarad$ can be expressed as
 \beq
 \label{finalzetar}
  {\cal P}_{\zetarad}={\cal P}_{\zeta_\i}+\frac{r^2}{9}{\cal P}_{S_G} \;,
 \eeq
where the entropy power spectrum amplitude is given by
 \beq
 {\cal P}_{S_G}=\frac{4}{\curv_*^2}\left(\frac{H_*}{2\pi }\right)^2
  \;.
 \eeq
In the case of
single field inflation
we have
 \beq
  {\cal P}_{\zeta_\i}=\frac{1}{2\mP^2\epsilon_*}\left(\frac{H_*}{2\pi }\right)^2 \;,
 \eeq
where $\epsilon_*\equiv -\dot H_*/H_*^2$ is the usual slow-roll
parameter during inflation
and $\mP^2=(8\pi G)^{-1}$ is the reduced Planck mass.
In order to compare the relative contributions of the inflaton and
of the curvaton in the final power spectrum (\ref{finalzetar}), it
is useful to introduce the dimensionless parameter
\cite{Vernizzi:2005fx}
 \beq
\label{lambda} \lambda \equiv \frac{8}{9}r^2\epsilon_*
\left(\frac{\mP}{\curv_*}\right)^2
 \eeq
so that ${\cal P}_{\zetarad}=(1+\lambda){\cal P}_{\zeta_\i}$.
If $\lambda\gg 1$, one recovers the standard curvaton scenario where
the inflaton perturbations can be ignored: since $r$ and $\epsilon_*$
are bounded by $1$, this requires $\curv_*\ll \mP$. A value of
$\lambda$ of order $1$ or smaller is possible if $r$ or $\epsilon_*$
are sufficiently small and/or $\curv_*$ is of the order of $\mP$. In
the present work, we will always assume $\curv_*\ll \mP$. If this is
not the case the curvaton starts to oscillate at about the same time
as it decays and cannot be described as a dust field (see
\cite{Langlois:2004nn} for details).

In slow-roll inflation the 3-point function of the inflaton
perturbations, $\zeta_\i$, is suppressed by slow-roll parameters
\cite{Maldacena:2002vr,Acquaviva:2002ud} and large
non-Gaussianities can arise only from the curvaton contribution.
Indeed, the 3-point function of $\zetarad$ yields (see also
\cite{Ichikawa:2008iq} for a similar analysis)
 \beq
  \langle \zetarad (\vec k_1) \zetarad(\vec k_2)
 \zetarad(\vec k_3) \rangle = (2 \pi)^3 \delta (\Sigma_i \vec
k_i) b^{\zeta\zeta\zeta}_{NL} \left[ P_{\zetarad}(k_1)
P_{\zetarad}(k_2) + {\rm
    perms} \right]\;, \label{3pointzeta}
 \eeq
where
$P_{\zetarad}(k)=2\pi^2 {\cal P}_{\zetarad}(k)/k^3$ and
$b^{\zeta \zeta\zeta}_{NL}$ is a non-linear parameter given in this
case by
 \beq
b^{\zeta \zeta\zeta}_{NL} =
\frac{1}{r}\frac{\left(\frac32-2r-r^2\right)}{(1+\lambda^{-1})^2}
\label{f_ad}\;,
 \eeq
as follows from Eq.~(\ref{zeta_rad3}). Non-Gaussianities are thus
significant when the curvaton decays well before it dominates,
$r\ll 1$.

When $\lambda\gg 1$ and the perturbations from inflation are
negligible, one recovers the standard curvaton result
\cite{Bartolo:2003bz}
and $b^{\zeta \zeta\zeta}_{NL}$ is proportional to the much used
local non-linear parameter $f_{NL}$ defined by $\zetarad=\zeta_G
+(3/5)f_{NL}\zeta_G^2$, i.e., $b^{\zeta \zeta\zeta}_{NL} =
(6/5)f_{NL}$. However, in general $b^{\zeta \zeta\zeta}_{NL}$ is
different from $f_{NL}$. Indeed, for other values of $\lambda$,
although only the curvaton contributes to the 3-point function,
the 2-point function depends also on the initial inflaton
fluctuation, which is a Gaussian random field,
independent of the curvaton fluctuation. This differs from the
original definition of $f_{NL}$ where only one Gaussian random
field is present \cite{Komatsu:2001rj}.

It is instructive to see how (\ref{f_ad}) depends on the curvaton
expectation value during inflation, $\curv_*$.
Substituting the relation $r\sim
(\curv_*/\mP)^2/\sqrt{\Gamma_\curv/m_\curv}$ (valid in the limit
$r\ll 1$) \cite{Gupta:2003jc}, where $\Gamma_\curv$ is the decay
rate of the curvaton, into the definition (\ref{lambda}), one sees
that $\lambda$ is proportional to $\curv_*^2$, like $r$. One then finds
that $b^{\zeta \zeta\zeta}_{NL}$ given in (\ref{f_ad}) reaches its maximal
value $b^{\zeta \zeta\zeta}_{NL}({\rm max})\sim \epsilon_*/\sqrt{\Gamma_\curv/m_\curv}$  for
$\lambda\sim 1$, i.e., for $\curv_*\sim
\sqrt{\Gamma_\curv/(m_\curv\epsilon_*) }\mP$. A significant non-Gaussianity is thus possible
if $\epsilon_*\gg \sqrt{\Gamma_\curv/m_\curv}$. Note also that when $r$ becomes small $b^{\zeta
\zeta\zeta}_{NL}$ does not grow indefinitely as one would naively
expect by considering $f_{NL} \simeq 5/(4r)$.
Finally, in the limit $r\ll 1$ and $\lambda\ll 1$, where the inflaton contribution dominates the power spectrum, the expression (\ref{f_ad}) simplifies into
 \beq
b^{\zeta \zeta\zeta}_{NL}\simeq \frac{3}{2} \frac{\lambda^2}{r}\sim
\frac{\epsilon_*^2
m_\curv^{3/2}}{\Gamma_\curv^{3/2}}\frac{\curv_*^2}{\mP^2}\;, \qquad
(\lambda\ll 1, \quad r\ll 1)\;. \label{rsmallNG}
\eeq

After the analysis of the non-Gaussianities for the {\it
adiabatic} perturbation, which essentially agrees with the
discussion given in \cite{Ichikawa:2008iq}, let us now turn to
entropy perturbations between CDM and radiation,
 \beq
  \label{defSrad}
 \Sm=3\left(\zetam-\zetar\right) \;,
 \eeq
which could be
generated {\it in the radiation era after the curvaton decay}.
If all the particle species are in full thermal equilibrium after
the curvaton decays, with vanishing chemical potentials, then the
primordial density perturbation must be adiabatic
\cite{Lyth:2002my,Weinberg:2004kf} and we have $\zetam=\zetar$ and
hence $\Sm=0$.
However, if CDM remains decoupled from (part of) the radiation, the
curvaton isocurvature perturbation may be converted into a
residual isocurvature perturbation after the curvaton decays.
We now consider two possibilities leading to a non-trivial isocurvature
perturbation \cite{Lyth:2002my}.

\subsection{CDM created {\em before} curvaton decay}

If the CDM is created before the
curvaton decay, then $\zetam=\zeta_\i$, which generates
 \beq
\Sm=3\left(\zeta_\i-\zetarad\right) =- r S_G-
\frac{r}{6}\left(\frac32-2r-r^2\right)S_G^2 \;.
 \eeq
This implies that the ratio between the isocurvature and adiabatic power spectra is given by
\beq
\frac{{\cal P}_{{\cal S}_{\rm m}} }{{\cal P}_{{\zeta}_{\rm r}}} = \frac{9}{1+\lambda^{-1}}\;.
\eeq
This quantity is constrained to be small by the CMB data.
In the case where the curvaton dominates the final $\zeta$, i.e. $\lambda\gg 1$, this
scenario is thus ruled out and this case is often disregarded in the literature \cite{Gordon:2002gv}.
However, if the inflaton contribution is sufficiently important,
$\lambda \ll 1$, such an entropy contribution is allowed. More
specifically,
the observational constraint
on $\alpha$, defined by ${\cal P}_{{\cal S}_{\rm m}}/{\cal
  P}_{\zeta_{\rm r}}\equiv \alpha/(1-\alpha)$,  is currently
$\alpha_0<0.067$ at $95 \%$ CL \cite{Komatsu:2008hk}. The subscript $0$ refers to the case where the
entropy and adiabatic fluctuations are {\it un-correlated}, which is appropriate here when $\lambda\ll 1$.
The non-Gaussianity of $\zeta$ is described by Eq.~(\ref{rsmallNG}).
Thus, it can become significant without
violating the current bound on the presence of isocurvature component
in the power spectrum.

The amount of non-Gaussianity in the temperature fluctuations of
the CMB anisotropies will depend both on $\zetarad$ and $\Sm$.
Thus, a complete study of these anisotropies would require the
knowledge of the 3-point correlation properties of both variables.
One can generalize Eq.~(\ref{3pointzeta}) and define
 \beq
\langle X (\vec k_1) Y (\vec k_2) Z(\vec
k_3) \rangle = (2 \pi)^3 \delta (\Sigma_i \vec k_i) b^{XYZ}_{NL}
\left[ P_{\zetarad}(k_1) P_{\zetarad}(k_2) + {\rm
    perms} \right]\;, \label{3pointall}
 \eeq
where $X,Y,Z$ can be $\zeta_{\rm r}$ or ${\cal S}_{\rm m}$. In this case all the
3-point functions have the same amplitude
(up to numerical factors
of $-3$), $b^{{\cal S} {\cal S}
{\cal S}}_{NL} = -3 b^{{\cal S} {\cal S} \zeta}_{NL} = 9 b^{{\cal
S} \zeta\zeta}_{NL} =  -27   b^{\zeta \zeta\zeta}_{NL}$, and
equally contribute to the non-Gaussianity of the CMB temperature
anisotropies.

\subsection{CDM created {\em from} curvaton decay}

The second possibility leading to non-trivial isocurvature
perturbation is when the local matter density is produced solely
from the local curvaton density (for instance, some fraction of the
curvaton decays to produce CDM particles or the out-of-equilibrium
curvaton decay generates the primordial baryon asymmetry). Then we
expect the matter density to be directly proportional to the
curvaton density on the decay hypersurface
 \begin{equation}
 \bar\rho_{\rm cdm} e^{3(\zetam-\zetar)} = c \bar\rho_\curv
  e^{3(\zeta_\curv-\zetar)} \,,
\end{equation}
where $c=(\bar\rho_m/\bar\rho_\curv)\ll1$, and hence to all orders
$\zetam = \zeta_\curv $. The matter isocurvature perturbation
(\ref{defSrad}) is then given by
 \begin{equation}
 \label{Smfromcurvaton}
 \Sm = 3(\zeta_\curv-\zetarad)=
{\cal S}_\curv+ 3( \zeta_\i-\zetarad)
 =(1-r)\left(S_G-\frac{3+6r+2r^2}{12}S_G^2\right) \,.
 \end{equation}
This implies that the ratio between the isocurvature and adiabatic power spectra  is
\beq
\frac{{\cal P}_{{\cal S}_{\rm m}} }{{\cal P}_{{\zeta}_{\rm r}}}
= \frac{9(1-r)^2}{r^2(1+\lambda^{-1})}\;.
\label{S_A}
\eeq
This quantity can be small, as required by  observations, in two limiting cases. Either $r$ is very close
to $1$ or $\lambda$ is very small.
In the pure curvaton model ($\lambda \gg 1$), $r$ is constrained to
be very close to $1$,
\beq
9(1-r)^2\simeq \alpha_{-1} < 0.0037 \quad (95 \% \ {\rm CL}),
\eeq
using the constraint given in \cite{Komatsu:2008hk}
for the totally {\em anti-correlated}
case,\footnote{Reference~\cite{Komatsu:2008hk} uses the same
  convention as ours for the sign of the adiabatic and isocurvature
  perturbations. However, the cross-correlation is defined with
  opposite sign to $\langle \zeta_{\rm r} {\cal S}_{\rm m}\rangle$, so that
in the pure curvaton case adiabatic and isocurvature perturbations are
referred to as being {\em anti-correlated}.}
and the non-linearity parameters $b^{XYZ}_{NL}$ defined in
Eq.~(\ref{3pointall}) involving entropy perturbations are suppressed by factors of $(1-r)$ with
respect to $b^{\zeta \zeta\zeta}_{NL}$, itself of order unity.

If the inflaton dominates the linear perturbations, i.e.
$\lambda\ll 1$, then the
ratio (\ref{S_A})  can be small even if the curvaton decays
long before it dominates.
For non-Gaussianity, the small $r$ limit
appears interesting
because the amplitudes of the 3-point correlation
functions $b^{XYZ}_{NL}$ are related by
\beq
b^{{\cal S} {\cal S} {\cal S}}_{NL} \sim  \frac{3}{r} \, b^{{\cal
S}
  {\cal S} \zeta}_{NL} \sim \frac{9}{r^2} \,
    b^{{\cal S}
  \zeta \zeta}_{NL} \sim \frac{27}{r^3} \,
   b^{\zeta \zeta
  \zeta}_{NL} \;,
\eeq
which shows that the amplitude of $b^{{\cal S} {\cal S} {\cal
S}}_{NL}$ can be much larger than that of $b^{\zeta \zeta
\zeta}_{NL}$ in this small $r$ limit.

However
this situation is viable only if $\lambda$ satisfies the constraint
 \beq
 \label{lambdaconstraint}
\frac{{\cal P}_{{\cal S}_{\rm m}} }{{\cal P}_{{\zeta}_{\rm r}}} \simeq \frac{9\lambda}{r^2} \simeq \alpha_{0} < 0.067 \quad (95 \% \ {\rm CL}) \qquad \left( r\ll 1,\quad \lambda \ll 1 \right),
 \eeq
using the constraint given in \cite{Komatsu:2008hk} for the {\it
un-correlated} case.
In this case $b^{\zeta \zeta \zeta}_{NL}$ is given by its limit in
Eq.~(\ref{rsmallNG}) and the expression for the
non-Gaussianity of the isocurvature component,
\beq
b^{{\cal S} {\cal S} {\cal S}}_{NL}=- \frac{27}{2}(3+6r+2r^2)\frac{(1-r)^3}{r^4(1+\lambda^{-1})^2}
\eeq
reduces to the simpler form
 \beq
  b^{{\cal S} {\cal S} {\cal S}}_{NL}
 \simeq - \frac{1}{2}\left( \frac{9\lambda}{r^2} \right)^2
 \simeq -32 \epsilon_*^2 \left( \frac{\mP}{\chi_*}\right)^4 \;,
 \eeq
where one recognizes the square of the isocurvature/adiabatic ratio ${\cal P}_{{\cal S}_{\rm m}}/{\cal P}_{{\zeta}_{\rm r}}$, which must be very  small because of the constraint (\ref{lambdaconstraint}).

Thus, even if the non-Gaussianity of the isocurvature component is much bigger than that of the adiabatic
component,  the constraint on the amplitude of the linear isocurvature
perturbations (\ref{lambdaconstraint}) also constrains the magnitude
of the bispectrum of isocurvature perturbations to be small.
This follows simply from the fact that in this case the second order
part of the matter isocurvature perturbation, $\Sm$ in
Eq.~(\ref{Smfromcurvaton}) is of order $\Sm^2$, and if the linear
part is constrained, then so is the non-linear part.

\section{Double quadratic inflation}

In this section, we consider double quadratic inflation
\cite{Polarski:1992dq}, i.e., an inflationary phase driven by two
massive and minimally coupled scalar fields described by the
Lagrangian
 \beq
{\cal L} = -  \frac{1}{2}(\partial_\mu \phi)^2 - \frac{1}{2}
(\partial_\mu \chi)^2 -V(\phi,\chi) \;, \qquad V(\phi,\chi)
=\frac{1}{2} \mp^2 \phi^2+ \frac{1}{2} \mc^2 \chi^2\;.
 \eeq

The adiabatic curvature perturbations generated by this model have
been computed in
\cite{Polarski:1992dq,Garcia-Bellido:1995kc,Mukhanov:1997fw} and
their non-Gaussianity in
\cite{Vernizzi:2006ve,Alabidi:2005qi,Rigopoulos:2005ae} (see also
\cite{Bernardeau:2002jy}). This model can also generate
isocurvature perturbations. At linear order these have been
computed in \cite{Polarski:1994rz} and it
was first noticed in \cite{Langlois:1999dw}
that they can be
correlated with the adiabatic ones. Here we extend these results
at second order and we compute the 3-point correlation properties
of the isocurvature perturbations and their correlation  with the
adiabatic perturbations.

We start by computing the adiabatic curvature perturbation $\zeta$
at second order, using the $\delta N$-formalism.
The expression of the number of e-folds in terms of the scalar
fields can be obtained from the slow-roll equations of motion,
which read
 \beq
3H^2\mP^2=\frac{1}{2}m_\phi^2 \phi^2+\frac{1}{2} \mc^2 \chi^2\,,
\quad 3H\dot\phi+\mp^2\phi=0\, , \quad 3H\dot\chi+\mc^2\chi=0\;, \label{sr}
 \eeq
and imply ${dN}/{dt}=H=-(\phi\dot\phi+\chi\dot\chi)/(2\mP^2)$.
For a given scale, the
number of e-folds between horizon crossing $t_*$
and some subsequent time $t$ is given
by the expression
 \beq
 N = \frac{1}{4 \mP^2} \left( \phi_*^2 +\chi_*^2 - \phi^2 -\chi^2 \right) \,,  \label{NNN}
 \eeq
where $\phi=\phi(t)$ and $\chi=\chi(t)$. Furthermore, from the last two
slow-roll equations
in (\ref{sr})
one can derive
 \beq
\left( \frac{\phi}{\phi_*} \right)^{R^2} = \frac{\chi}{\chi_*} \,,
\label{Relation_C}
 \eeq
where
$R\equiv {\mc}/{\mp}$
and without loss of generality we take
$R\geq1$.

To compute $\zeta$,
we choose as final hypersurface at time $t$
a uniform density hypersurface defined by the condition
 \beq
  R^2 \chi^2 +\phi^2=C
  \;.
  \label{uniform_density_C}
 \eeq
Then, Eqs.~(\ref{Relation_C})
and (\ref{uniform_density_C}) uniquely
fix the relation between $(\phi,\chi)$ and $(\phi_*,\chi_*)$.
Indeed, by combining these two equations we can derive
 \beq
  R^2 \left(\frac{\phi}{\phi_*} \right)^{{2}{R^2}}
\chi_*^2 +\phi^2 =C, \qquad R^2 \chi^2 +\left(\frac{\chi}{\chi_*}
\right)^{\frac{2}{R^2}} \phi_*^2 = C \;,
 \eeq
which can be used to find the derivatives of $\phi$ and $\chi$
with respect to $\phi_*$ and $\chi_*$. These relations can then be
employed to compute the first and second derivatives of $N$ with
respect to $\phi_*$ and $\chi_*$ by differentiating
Eq.~(\ref{NNN}). The calculation is reported in Appendix
\ref{app:derivatives}. By using Eq.~(\ref{zeta_N}) up to second
order in $\delta \phi_*$ and $\delta \chi_*$, one obtains an
expression for $\zeta$,
 \bea
\zeta &=& \frac{1}{2 \mP^2} \left( \bar \phi_* \delta \phi_* +
\bar \chi_* \delta \chi_*  + \frac{1}{2}\delta \phi_*^2 +
\frac{1}{2}\delta
\chi_*^2 \right) \nonumber \\
 &&+ \frac{(R^2-1)}{2 \mP^2}\bar g \left[ \frac{\delta
 \chi_*}{\bar \chi_*} \left( 1-\frac{\delta \chi_*}{2\bar\chi_*} \right) -
 R^2\frac{\delta \phi_*}{\bar\phi_*} \left(1-\frac{\delta
 \phi_*}{2\bar\phi_*} \right) +\frac{\bar\chi_*\bar g_{,\chi_*}}{2\bar g}
 \left(\frac{\delta \chi_*}{\bar \chi_*} - R^2\frac{\delta
 \phi_*}{\bar \phi_*} \right)^2 \right] \,,
  \label{zetaintime}
 \eea
 where $g=g(\phi,\chi)$ is defined as $ g= \phi^2 \chi^2/(R^4
\chi^2 +\phi^2)$.
This relation holds until the end of slow-roll inflation.

We will consider the following scenario. Inflation is initially
driven by the heavy field, $\chi$, which slow-rolls down the
potential. Later, the heavy field becomes subdominant and then
starts oscillating, while the light field, $\phi$, drives
inflation.
In the last stage of slow-roll inflation, when $\bar
\chi\ll \mP$, the coefficients $\bar g/\mP^2$ and $\bar\chi_*\bar
g_{,\chi_*}/\mP^2$ are very small and the second line of
Eq.~(\ref{zetaintime}) becomes negligible. The curvature
perturbations $\zeta$ thus becomes effectively constant, and since
its value is unaffected by the subsequent stages of inflation and
reheating, one finds
the expression
 \beq
 \zetarad = \frac{1}{2 \mP^2} \left(
\bar \phi_* \delta \phi_* + \bar \chi_* \delta \chi_* +
\frac{1}{2}\delta \phi_*^2 + \frac{1}{2}\delta \chi_*^2  \right)
\;\label{zetarad}
 \eeq
for the adiabatic perturbation during the radiation era.

Let us now focus on the isocurvature perturbation which can be
produced, after inflation, in this type of model. To determine it,
it is convenient to use the relative comoving curvature
perturbations ${\cal R}_\phi$ and ${\cal R}_\chi$. According to
Eq.~(\ref{zeta_field}), they are given by
 \beq
{\cal R}_{\phi} = \delta N - \int_{\bar \phi}^{\phi} H
\frac{d\tilde \phi}{ \dot{ \tilde \phi}}\;, \qquad  {\cal
R}_{\chi} = \delta N - \int_{\bar \chi}^{\chi} H \frac{d \tilde
\chi}{ \dot{\tilde \chi}}\;. \label{def_zeta_chi}
 \eeq

The light field, $\phi$, remains in slow-roll all the time during
inflation.\footnote{We assume that there is no intermediate
dust-like phase between the heavy field dominated inflation and
the light field dominated inflation.} In order to compute ${\cal
R}_\phi$ at second order we can use Eq.~(\ref{zetaI_N}) for
$\varphi_A = \phi$, and expand $N^{(\phi)}$, i.e., the number of
e-folds from an initial flat hypersurface at $t_*$ to a final
uniform field $\phi$ hypersurface, up to second order in $\delta
\phi_*$ and $\delta \chi_*$. To compute $N^{(\phi)}$ we substitute
Eq.~(\ref{Relation_C}) in Eq.~(\ref{NNN}) and impose that the
final value of $\phi$ is a constant, $\phi=C_\phi$. Using this
condition and differentiating $N^{(\phi)}$ with respect to the
initial field values (see Appendix \ref{app:derivatives}),
Eq.~(\ref{zetaI_N}) then yields
 \beq
{\cal R}_{\phi} = \zeta_r + \frac{\bar \chi^2}{2 \mP^2} \left( R^2
\frac{\delta \phi_*}{\bar\phi_*} - \frac{\delta \chi_*}{\bar
\chi_*} -\frac{2R^4+R^2}{2}\frac{\delta \phi_*^2}{\bar \phi_*^2} -
\frac{1}{2} \frac{\delta \chi_*^2}{\bar \chi_*^2} +2R^2
\frac{\delta \phi_*}{\bar \phi_*}\frac{\delta \chi_*}{\bar \chi_*}
\right) \;, \label{Rphi}
 \eeq
where the explicit expression for $\zeta_r$ is given in
(\ref{zetarad}). At the end of inflation $\phi$ dominates and
reheats the universe. When $\phi$ becomes dominant its comoving
and uniform energy density curvature perturbations are the same on
large scales, $\zeta_\phi = {\cal R}_\phi$.
Furthermore, when $\bar \chi^2\ll\mP^2$, the second term on the
right hand side of Eq.~(\ref{Rphi}) becomes negligible, and ${\cal
R}_\phi=\zeta_{\rm r}$.

The evolution of the perturbation of the heavy field $\chi$ is
more complicated. During the $\phi$-dominated
slow-roll phase, a calculation similar to the
one for $\phi$ yields, replacing $\phi$ by $\chi$ and $R^2$ by
$R^{-2}$ in Eq.~(\ref{Rphi}),
\bea
{\cal R}_\chi|_{\rm slow-roll} = \zeta_r+ \frac{3H^2}{m_\phi^2}
\left( \frac{1}{R^2}\frac{\delta \chi_*}{\bar \chi_*} -
\frac{\delta \phi_*}{ \bar \phi_*} -
\frac{2+R^2}{2R^4}\frac{\delta \chi_*^2}{\bar \chi_*^2} -
\frac{1}{2} \frac{\delta \phi_*^2}{\bar \phi_*^2} +\frac{2}{R^2}
\frac{\delta \phi_*}{\bar \phi_*}\frac{\delta \chi_*}{\bar \chi_*}
\right) \;,
\label{zeta_chi_sr}
\eea
where we have used that $\phi$ dominates the Universe and thus
$H^2= {m_\phi^2 \bar \phi^2}/({ 6 \mP^2})$. This expression is
valid when $\chi$ is subdominant and in slow-roll but cannot be
used when $\chi$ oscillates. It is convenient to use
Eq.~(\ref{def_zeta_chi}) to rewrite Eq.~(\ref{zeta_chi_sr})  in
terms of the field fluctuation $\delta \chi$ on a constant total
energy density hypersurface ($\delta N = \zeta$),
\beq
{\cal R}_\chi|_{\rm slow-roll} = \zeta -\int_{\bar \chi}^{\chi} H
\frac{d \tilde \chi}{ \dot{\tilde \chi}} = \zeta+ \frac{3H^2}{m_\chi^2}
\int_{\bar \chi}^{\chi} \frac{d\tilde \chi}{ \tilde \chi} \;,
\eeq
where we have used the property that $H$, which depends only on
the slow-rolling $\phi$, is (spatially) constant on a constant
total energy density hypersurface, since the latter coincides with
a constant $\phi$ hypersurface when $\chi$ is subdominant.
This yields the non-linear relation between the isocurvature
perturbation during slow-roll and the local value of the heavy
field,
\beq
\chi = \bar \chi e^{\frac{m_\chi^2}{3H^2} ({\cal R}_\chi|_{\rm
slow-roll} - \zeta)}\;,
\eeq
which expanded up to second order yields
\beq
{\cal R}_\chi|_{\rm slow-roll} = \zeta+\frac{3H^2}{m_\chi^2}\left(
\frac{\delta \chi}{\bar \chi} - \frac12 \frac{\delta \chi^2}{\bar
\chi^2} \right)\;. \label{iso_sr}
\eeq

As for the curvaton, when $\chi$ oscillates we can describe it as
a non-relativistic fluid and use Eq.~(\ref{rho_curvaton}).
Expanding this equation up to second order in the field
fluctuation $\delta \chi$, we obtain the curvature perturbation
$\zeta_\chi$ during the oscillations,
\beq
\zeta_\chi|_{\rm osc} =
\zeta_{\rm r}+ \frac{2}{3}\left(
\frac{\delta \chi}{\bar \chi} - \frac12 \frac{\delta \chi^2}{\bar
\chi^2} \right)\;. \label{iso_osc}
\eeq
Now we can use the constancy of $\delta \chi /\bar \chi$ valid up
to second order to match Eqs.~(\ref{iso_sr}) and (\ref{iso_osc})
and
express $\zeta_\chi|_{\rm osc}$ in terms of ${\cal R}_\chi|_{\rm slow-roll}$. Using Eq.~(\ref{zeta_chi_sr}),
 we find that the value of $\zeta_\chi$ after  inflation is
 \beq
 \zeta_\chi= \zetarad +\frac23
 R^2 \left( \frac{1}{R^2}\frac{\delta \chi_*}{\bar \chi_*} - \frac{\delta
\phi_*}{ \bar \phi_*} -
\frac{2+R^2}{2R^4}\frac{\delta \chi_*^2}{\bar \chi_*^2}
- \frac{1}{2} \frac{\delta \phi_*^2}{\bar \phi_*^2} +\frac{2}{R^2}
\frac{\delta \phi_*}{\bar \phi_*}\frac{\delta \chi_*}{\bar \chi_*}
\right)
 \;.
 \eeq

We assume that the light field decays into radiation which
dominates the Universe after inflation, and that the heavy field
decays into CDM when it oscillates. Then \beq \Sm =3( \zeta_{\chi}
-\zetarad) \;, \label{S_osc} \eeq
and
\beq
\Sm= 2 R^2
\left( \frac{1}{R^2}\frac{\delta \chi_*}{\bar \chi_*} - \frac{\delta
\phi_*}{ \bar \phi_*} -
\frac{2+R^2}{2R^4}\frac{\delta \chi_*^2}{\bar \chi_*^2}
- \frac{1}{2} \frac{\delta \phi_*^2}{\bar \phi_*^2} +\frac{2}{R^2}
\frac{\delta \phi_*}{\bar \phi_*}\frac{\delta \chi_*}{\bar \chi_*}
\right)
\;. \label{Scdm}
\eeq
This equation generalizes at second order the results of
\cite{Polarski:1994rz}.

At this point it is useful to express the final curvature and
entropy perturbations in terms of the instantaneous adiabatic and
entropy perturbations during inflation (more precisely when the
scale of interest exits the Hubble radius). The decomposition of
two scalar field perturbations in terms of (instantaneous)
adiabatic and entropy  perturbations has been introduced at linear
order in \cite{Gordon:2000hv} and generalized at non-linear order
in \cite{Langlois:2006vv}.  The general definitions are recalled
in Appendix \ref{app:decomposition}. Here we give the expressions
for the adiabatic perturbation $\delta \sigma$ and the entropy
perturbation $\delta s$ for the particular case of double
quadratic inflation.
To simplify the notation, let us define
 \beq
\co=\cos\theta=-\bar \phi/\x\;, \qquad \si=\sin\theta=-R^2\bar
\chi/\x\;,
 \eeq
with $\x=(\bar \phi^2  + R^4 \bar\chi^2)^{1/2}$. The angle
$\theta$ is simply the angle between the instantaneous direction
of the field trajectory and the $\phi$-axis.
At first order in perturbations we have
 \beq
\delta \sigma^{(1)} = \co \delta \phi+\si \delta \chi \;, \qquad
\delta s^{(1)} = \co \delta \chi-\si \delta \phi \;,
\eeq
while the second order expression are given by
\bea
\delta \sigma &=& \delta \sigma^{(1)} -
\frac{R^2\co^2+\si^2}{2\x}
\delta s^{(1)} \delta s^{(1)}
 \;, \\
\delta s &=&
 \delta s^{(1)} +
\frac{R^2\co^2+\si^2}{\x}
\delta s^{(1)} \delta \sigma^{(1)}
+ \frac{(R^2-1)\co\si}{2\x}
 \delta \sigma^{(1)}{}^2
 \;.
\eea
Note that at second order the definition of $\delta \sigma$
contains first order perturbations of $\delta s^{(1)}$
and vice verse. Indeed, as explained in \cite{Langlois:2006vv},
the adiabatic and
entropy field decomposition is local and second order fluctuations
will be sensitive to first order fluctuations of the angle $\theta$,
which can be re-expressed in terms of the field fluctuations $\delta
\sigma^{(1)}$ and $\delta s^{(1)}$.

Using these definitions and evaluating Eq.~(\ref{zetaintime}) at $t=t_*$,
we can rewrite $\zeta_*$, i.e. the curvature perturbation on uniform
density hypersurfaces at Hubble crossing,
\bea
\zeta_* &=& - \frac{1}{2 \mP^2} \left[  \left(\co_*^2+R^{-2}\si_*^2\right) \xi_*\delta \sigma_*
- \frac{1}{2} \left(
  1- R^{-2}(1-R^2)^2\co_*^2\si_*^2 \right) \delta \sigma_*^2
  \right. \nonumber \\
&&+ \left.
      (R^2-1)\co_*\si_*(\co_*^2+R^{-2}\si_*^2)
\delta \sigma_* \delta s_*
\right]
\;. \label{zetastar}
\eea
The second order expression for $\zeta_*$ contains also the first
order entropy field perturbation. Indeed, this is the case also
for its general slow-roll form (\ref{zetastar-general}) given in
Appendix \ref{app:decomposition}, derived in
\cite{Langlois:2006vv}.

The entropy field perturbation sources $\zeta$ at first and second
order (see Eq.~(\ref{zetadot}) in Appendix
\ref{app:decomposition}). Using Eqs.~(\ref{zetarad}) and
(\ref{zetastar}), the final value of $\zeta$ is thus
 \beq
  \zetarad = \zeta_*+ \frac{1-R^{-2}}{2 \mP^2}
\xi_* \co_* \si_* \left(\delta s_* - \frac{R^2\co_*^4
-\si_*^4}{2 \co_* \si_* \xi_* } \delta s_*^2 \right)  \;.
 \label{zetarad2fields}
 \eeq
Furthermore, we can rewrite the expression for $\Sm$,
Eq.~(\ref{Scdm}), in terms of the adiabatic and entropy field
perturbations. This reads
%
 \beq
  \Sm = -\frac{2 R^2}{\co_* \si_* \xi_*} \left( \delta s_* +
\frac{1+2\co_*^2 +( R^2-1) \co_*^4 }{2 \co_* \si_* \xi_* } \delta
s_*^2
  \right) \;. \label{Scdm2fields}
\eeq
Neglecting slow-roll corrections, the field perturbations $\delta
\sigma_*$ and $\delta s_*$ are random fields with 2-point
functions given by
\beq
\langle \delta \sigma_*(\vec k) \delta \sigma_*(\vec k') \rangle = \langle
\delta s_*(\vec k) \delta s_*(\vec k') \rangle =
(2\pi)^3 \delta (\vec k+\vec k')\frac{H^2_*}{2 k^3}  \;, \qquad \langle
\delta \sigma_*(\vec k) \delta s_*(\vec k') \rangle=0\;.
\eeq
At lowest order in slow-roll, these fields are Gaussian. However,
their 3-point correlation functions are non vanishing and have
been computed in Appendix \ref{app:decomposition}. From the
expression of $\zeta_*$, Eq.~(\ref{zetastar}), and from the
general definition of $\zeta$ given in Appendix
\ref{app:decomposition}, Eq.~(\ref{zetastar-general}), one can see
that there are second order corrections proportional to $\delta
\sigma_*^2$ and $\delta \sigma_* \delta s_*$ so that $\zeta_*$ has
not exactly the same correlation properties as $\delta \sigma_*$.
However, these contributions are {\em generically} (for any
slow-roll model) of the same order in slow-roll as the 3-point
function of $\delta \sigma_*$ so that at lowest order in slow-roll
$\zeta_* \propto \delta \sigma_*$ is a Gaussian random field.

A more convenient parametrization often used in the literature
\cite{Polarski:1992dq} is to rewrite the background values of the
scalar fields in polar coordinates,
\beq
\bar \phi = 2\mP \sqrt{N_{\rm e} -N} \cos \alpha\;, \qquad \bar
\chi = 2\mP \sqrt{N_{\rm e} -N} \sin \alpha\;.
\eeq
In terms of the angle $\alpha$ and of $N_{\rm e}-N_*$, the number
of e-folds from Hubble crossing to the end of inflation, the power
spectrum of $\zeta_*$ is, using the linear term in
Eq.~(\ref{zetastar}),
\beq
{\cal P}_{\zeta_*} = (N_{\rm e} -N_*)  \frac{(1+R^2 \tan^2
\alpha_*)^2}{(1+\tan^2 \alpha_*)(1+R^4 \tan^2 \alpha_*)}
\left(\frac{H_*}{2\pi \mP}\right)^2\;.
\eeq
Furthermore, instead of using $\delta s_*$,
we find it useful to rewrite Eqs.~(\ref{zetarad2fields})
and (\ref{Scdm2fields}) in terms of $S_*$, defined as having the same
power spectrum of $\zeta_*$, i.e.,
 \beq
  S_*
\equiv
- \frac{1}{2 \mP^2} (\co_*^2+R^{-2}\si_*^2) \xi_* \delta
  s_*\;,
\qquad {\cal P}_{S_*} = {\cal P}_{\zeta_*} \;,
 \eeq
in analogy with the linear term of Eq.~(\ref{zetastar}). At
leading order in slow-roll $S_*$ is a Gaussian random field
uncorrelated with $\zeta_*$. Finally,
using this parametrization we obtain
 \beq
 \zetarad = \zeta_* + \frac{ (1-R^2)  \tan \alpha_*
}{1 + R^{2} \tan^2 \alpha_*} \left[ S_* + \frac{\eta_{\phi
\phi}}{2} \frac{1 - R^6\tan^4 \alpha_*}{\tan \alpha_* (1+R^{4}
\tan^2 \alpha_*)} S_*^2 \right] \;, \label{zetarad2fieldsalpha}
\eeq
and
\beq \Sm = 2 \eta_{\phi \phi} \frac{1+ R^4 \tan^2 \alpha_*}{
\tan \alpha_*} \left[ S_* - \frac{\eta_{\phi \phi}}{2} \frac{ 2 +
R^2 +4R^4\tan^2 \alpha_* +R^8 \tan^4 \alpha_*}{ R^2 \tan
\alpha_*(1+R^{4} \tan^2 \alpha_*)} S_*^2
\right] \;, \label{Scdm2fieldsalpha}
\eeq
where $\eta_{\phi \phi}$ is a slow-roll parameter,
\beq
\eta_{\phi \phi} =\frac{1+\tan^2 \alpha_*}{2 (N_{\rm e} -N_*)
(1+R^2 \tan^2 \alpha_*)} \;.
\eeq

As discussed in \cite{Langlois:1999dw}, adiabatic and entropy
perturbations can be correlated at linear order but the
correlation can be neglected when $ R^2 \tan \alpha_* \ll 1$ or
$\tan \alpha_* \gg 1$. Indeed, in this case
Eqs.~(\ref{zetarad2fieldsalpha}) and (\ref{Scdm2fieldsalpha})
yields, at linear order, $\zeta_{\rm r} = \zeta_*$ and ${\cal
S}_{\rm m} \propto S_*$. However, these equations show that
adiabatic and entropy perturbations are always correlated at
second order, even when they are uncorrelated at linear order. In
this particular model of two quadratic potential that we could
treat analytically, non-linear terms are small, being suppressed
by slow-roll. This leads to small non-Gaussianities in the
adiabatic perturbation (cf.~Ref.~\cite{Vernizzi:2006ve}) and also
in the entropy perturbation. However, we do not expect this to be
a generic feature of all inflation models. In particular, the
coefficients in front of the $S_*^2$ terms in
Eqs.~(\ref{zetarad2fieldsalpha}) and (\ref{Scdm2fieldsalpha}) may
be much larger in other models, which can lead to a non-vanishing
3-point correlation between the adiabatic and entropy
perturbations.

\section{Conclusions}

We have calculated the second-order primordial curvature and
isocurvature perturbations from two models of inflation in the early
universe. In the first example of a mixed curvaton-inflaton model we
assume the curvaton is an isocurvature field completely decoupled
from the inflaton field driving inflation. In the second, double
quadratic inflation model the two massive fields driving inflation
are gravitationally coupled during slow-roll.

The field perturbations at Hubble-exit during slow-roll inflation
are effectively independent Gaussian random fields; their
cross-correlation and non-linearities are suppressed by slow-roll
parameters. However the coupled evolution on large scales after
Hubble-exit can lead to cross-correlations at linear order
\cite{Langlois:1999dw,Gordon:2000hv} and we have calculated the
correlations that arise at second-order. This can lead to
non-vanishing bispectra for the primordial curvature and
isocurvature perturbations and their cross-correlations.

In both cases we find that the non-linear primordial curvature and
isocurvature perturbations (\ref{zeta_N}) and (\ref{Sm_N}) can be
given in terms of the adiabatic and entropy field perturbations at
horizon-exit during inflation,
\begin{eqnarray}
 \label{zetarfinal}
 \zetar &=& N_{,\sigma}\delta\sigma_* + N_{,s}\delta s_*
  + \frac12 N_{,ss}\delta s_*^2
  + \left[ \frac12 N_{,\sigma\sigma}\delta\sigma_*^2
  + N_{,s\sigma} \delta s_* \delta\sigma_* \right]
  \,, \\
  \label{Smfinal}
 \frac{1}{3} \Sm &=& \Delta N_{,s}\delta s_*
  + \frac12 \Delta N_{,ss}\delta s_*^2
  \,,
  \end{eqnarray}
where $N$ describes the expansion to a surface of uniform
radiation density in the radiation dominated era and $\Delta N$
describes the difference between the expansion to hypersurfaces of
uniform radiation density and uniform matter density. Note that
$\Delta N$ vanishes for adiabatic perturbations, i.e., when
$\delta s_*=0$.

In the mixed curvaton-inflaton model we identify the inflaton
field with the adiabatic perturbations during inflation,
$\delta\sigma_*$, and the curvaton field with entropy field
perturbations, $\delta s_*$. The bracketed terms in
Eq.~(\ref{zetarfinal})
%
can be neglected at leading order in a slow-roll approximation. It
is well known that $\zeta$ can have a significant non-Gaussianity
when $r\ll 1$ in the curvaton scenario (i.e., when  $\lambda$, the
ratio between curvaton and inflaton contributions to the curvature
power spectrum, is large).
However, in this case the primordial isocurvature
perturbations are constrained to be very small
\cite{Gordon:2002gv,beltran}. We have shown that
it is possible for a residual isocurvature perturbation to have a
bispectrum which is much larger than that of the adiabatic
perturbation if $\lambda$ is small, i.e., if the inflaton
perturbation dominates the primordial curvature perturbation at
first order,
and if the CDM is produced by the curvaton decay. However
observational constraints on the power spectrum of isocurvature
perturbations also constrains the
bispectra
to be small in this case.
The most interesting situation is the scenario where the CDM is created
before the curvaton decay, which is viable  if the inflaton contribution dominates
the linear power spectrum. In this case, it is possible to obtain a strong non-Gaussianity
if $\epsilon_*\gg \sqrt{\Gamma_\curv/m_\curv}$ and we have found that the non-Gaussianity of the adiabatic
component and of the isocurvature component are of the same order of magnitude.

In double quadratic inflation the two canonical fields, $\phi$ and
$\chi$, are coupled gravitationally and the adiabatic and entropy
field perturbations, $\delta\sigma_*$ and $\delta s_*$, are in
general a linear combination of the two canonical fields. We have
obtained explicit expressions at second order relating the initial
and final adiabatic and isocurvature perturbations. In this simple
case of two uncoupled fields with quadratic potentials we find
that the non-linearities are small, but this need not be the case
in other models. Indeed, as shown in Appendix
\ref{app:decomposition}, in general two field slow-roll inflation
we expect that the terms in square brackets in
Eq.~(\ref{zetarfinal})
%
can be neglected at leading order in slow-roll. However, the
remaining non-linear terms due to initial entropy perturbation
need not be suppressed. It would be interesting to investigate
non-Gaussianity of isocurvature perturbations in more general
models.

The bispectra for primordial curvature and isocurvature
perturbations and their cross-correlations are presented for a
general two-field model in Appendix \ref{app:primordial}, and
given at leading order in a slow-roll expansion.
These show that the non-zero primordial bispectra (in both curvature
and isocurvature perturbations and their cross-correlations) arise
due to entropy field perturbations at Hubble-exit. Our results for
the primordial curvature perturbation are consistent with the
non-linear $\delta N$-formalism \cite{Lyth:2005fi}, derived in the
large scale limit where the separate universes approach
\cite{Wands:2000dp} is used to evaluate the perturbed expansion
using the homogeneous background solutions. In single-field
slow-roll inflation the perturbations are adiabatic on large scales
and the bispectrum is suppressed by slow-roll parameters
\cite{Maldacena:2002vr,Acquaviva:2002ud}.

In multiple-field inflation there is the possibility of additional
observational features which are absent in single-field models. We
have shown that this could include the contribution of isocurvature
field perturbations to the bispectra (3-point functions) as well as
the power spectra (2-point functions). It is interesting to note
that, in principle, the isocurvature perturbations might dominate
the primordial bispectrum in, for example, the CMB temperature
anisotropies while remaining sub-dominant in the power spectrum. The
non-Gaussian primordial perturbations predicted from Gaussian field
perturbations during inflation are of a specific local form, but
should be distinguishable from the local non-Gaussianity of the
primordial curvature described by conventional $f_{NL}$ parameter.
The bispectrum of the isocurvature perturbations can be
characterized by a new non-linearity parameter, and the
cross-correlated bispectra yield additional parameters. But in
curvaton models, for example, they are all determined by the single
model parameter, $r$, and thus could provide a strong test of the
curvaton scenario. The best constraints on specific models of
non-Gaussianity are based on matched filtering techniques
\cite{Bartolo:2004if}. It will thus be important to develop
optimized constraints for the non-Gaussianity of primordial
isocurvature perturbations to obtain the optimal constraints on a
wider range of theoretical models.

\section*{Acknowledgments}
We thank the  organizers of the workshop ``Astroparticle and Cosmology''   at
the Galileo Galilei Institute for Theoretical Physics, Florence, October 2006, where this work was initiated, and
of the workshop ``Non-Gaussianity from fundamental physics'' at DAMTP, Cambridge, September 2008, where this work
was completed, for their kind invitation.
DL thanks the ICG for their hospitality.
FV acknowledges the kind hospitality of APC, Paris, and ICG,
Portsmouth, where part of this work was carried out.
DW is supported by the STFC.

\appendix

\section{Derivatives of $N$, $N^{(\phi)}$ and $N^{(\chi)}$ in double inflation}
\label{app:derivatives}

The total number of e-folds in double inflation is given by
 \beq
 N = \frac{1}{4 \mP^2} \left( \phi_*^2 +\chi_*^2 - \phi^2 -\chi^2 \right)
 \,. \label{Napp}
 \eeq
If $\phi$ and $\chi$ are the values of the fields on a final
uniform total density hypersurface, $R^2\chi^2+\phi^2=C$, using
Eq.~(\ref{Relation_C}) we obtain
 \beq
  R^2 \left(\frac{\phi}{\phi_*} \right)^{{2}{R^2}}
\chi_*^2 +\phi^2 =C, \qquad R^2 \chi^2 +\left(\frac{\chi}{\chi_*}
\right)^{\frac{2}{R^2}} \phi_*^2 = C \;,
 \eeq
which can be differentiate with respect to $\phi_*$ and $\chi_*$
to yield
\beq
\frac{\partial \phi}{\partial\phi_*} = \frac{R^4 }{\phi_* \phi}
g\;, \qquad \frac{\partial \phi}{\partial\chi_*} =  - \frac{R^2
}{\chi_*\phi} g \;,  \qquad \frac{\partial \chi}{\partial\phi_*} =
- \frac{R^2}{\phi_* \chi }g \;, \qquad \frac{\partial
\chi}{\partial\chi_*} = \frac{1}{\chi_*\chi} g \;,
 \eeq
where $ g= \phi^2 \chi^2/(R^4 \chi^2 +\phi^2)$. By differentiating
$N$ in Eq.~(\ref{Napp}) we get
\bea N_{,\phi_*} &=&
\frac{ \phi_*}{2 \mP^2} \left[1 +(1-R^2)R^2
\frac{g}{\phi_*^2} \right], \\
N_{,\chi_*} &=&  \frac{\chi_*}{2 \mP^2} \left[1 +(R^2-1)
\frac{g}{\chi_*^2} \right]\;,
 \eea
and
 \bea
N_{,\phi_* \phi_*}  &=&  \frac{1}{2 \mP^2} \left[ 1+ (1-R^2)R^2
\left(\frac{g_{,\phi_*}}{\phi_*} - \frac{g}{\phi_*^2} \right)\right]\;, \\
N_{,\phi_* \chi_*}  &=&  \frac{1}{2 \mP^2} (1-R^2)R^2
\frac{g_{,\chi_*}}{\phi_*} = N_{,\chi_* \phi_*} = \frac{1}{2
\mP^2}
(R^2-1) \frac{g_{,\phi_*}}{\chi_*}\;, \\
N_{,\chi_*\chi_*}  &=&  \frac{1}{2 \mP^2} \left[ 1+ (R^2- 1)
\left(\frac{g_{,\chi_*}}{\chi_*} - \frac{g}{\chi_*^2}
\right)\right]\;.
 \eea

If the final hypersurface is a uniform field $\phi$ hypersurface,
$\phi=C_\phi$, the number of e-folds (\ref{Napp}) reads,
 \beq
 N^{(\phi)} = \frac{1}{4 \mP^2} \left[
 \phi_*^2 +\chi_*^2 - C_\phi^2 -\left(\frac{C_\phi}{\phi_*}\right)^{2R^2}\chi_*^2 \right] \,, \eeq
which can be differentiated to give
\bea
N_{,\phi_*}^{(\phi)} &=&\frac{\phi_*}{2\mP^2}  \left(1 + R^2
\frac{\chi^2}{\phi_*^2} \right) \;, \\
N_{,\chi_*}^{(\phi)} &=&\frac{\chi_*}{2\mP^2}  \left(1 -
\frac{\chi^2}{\chi_*^2} \right) \;,
 \eea
and
 \bea
N_{,\phi_* \phi_*}^{(\phi)} &=&\frac{1}{2\mP^2} \left(1 -
R^2 \frac{\chi^2}{\phi_*^2} (1+2R^2) \right) \;, \\
N_{,\chi_* \phi_*}^{(\phi)} &=& N_{,\phi_* \chi_*}^{(\phi)}
=\frac{ R^2}{\mP^2}
\frac{\chi^2}{\phi_*\chi_*} \;, \\
N_{,\chi_*\chi_*}^{(\phi)} &=&\frac{1}{2\mP^2} \left(1 -
\frac{\chi^2}{\chi_*^2} \right) \;,
 \eea

Similar expressions can be found if we consider a final uniform
field $\chi$ hypersurface. In this case
\bea
N_{,\chi_*}^{(\chi)} &=&\frac{\chi_*}{2\mP^2}  \left(1 + R^{-2}
\frac{\phi^2}{\chi_*^2} \right) \;, \\
N_{,\phi_*}^{(\chi)} &=&\frac{\phi_*}{2\mP^2}  \left(1 -
\frac{\phi^2}{\phi_*^2} \right) \;,
 \eea
and
 \bea
N_{,\chi_* \chi_*}^{(\chi)} &=&\frac{1}{2\mP^2} \left(1 -
R^{-2} \frac{\phi^2}{\chi_*^2} (1+2R^{-2}) \right) \;, \\
N_{,\phi_* \chi_*}^{(\chi)} &=& N_{,\chi_* \phi_*}^{(\chi)}
=\frac{ R^{-2}}{\mP^2}
\frac{\phi^2}{\chi_*\phi_*} \;, \\
N_{,\phi_*\phi_*}^{(\chi)} &=&\frac{1}{2\mP^2} \left(1 -
\frac{\phi^2}{\phi_*^2} \right) \;.
 \eea

\section{Adiabatic and entropy field decomposition}
\label{app:decomposition}

In this section we review the adiabatic and entropy decomposition
approach at linear and second order, during slow-roll inflation.
It is possible to make a rotation in field space to identify the
instantaneous adiabatic and entropy field perturbations along and
orthogonal to the field trajectory. We will use the results
 of
\cite{Langlois:2006vv} generalizing the work of
\cite{Gordon:2000hv} (see \cite{NLdec} for an equivalent
approach). At linear order, the adiabatic and entropy field
perturbations are defined, respectively, as
 \beq
\delta \sigma^{(1)} = \cos \theta \delta \phi + \sin \theta \delta
\chi  \;, \qquad \delta s^{(1)} = -\sin \theta \delta \phi + \cos
\theta \delta \chi \;, \label{defrotlin}
 \eeq
where $\tan \theta =\dot \chi/\dot \phi$ and $\theta$ is the
time-dependent angle of the instantaneous rotation. At second order, we
define
 \beq
\delta \sigma = \delta \sigma^{(1)} + \frac{\delta s \dot{\delta
s}}{2 \dot \sigma} \;, \qquad \delta s = \delta s^{(1)} -
\frac{\delta \sigma}{\dot \sigma} \left(\dot {\delta s}
+\frac{\dot \theta}{2} \delta \sigma \right) \;, \label{defrotsec}
 \eeq
where $\dot \sigma^2=\dot \phi^2 +\dot \chi^2 = -2 \dot H \mP^2$.

The adiabatic curvature perturbation on uniform density
hypersurfaces is defined at second order as
 \beq
\zeta = -\frac{H}{\dot \sigma} \delta \sigma - \frac{\delta
  \sigma}{\dot \sigma} \left[ - \Tdot{\left(\frac{H}{\dot \sigma}
\delta \sigma \right)} + \frac{1}{2} \Tdot{\left(\frac{H}{\dot
\sigma} \right)} \delta \sigma + \dot \theta  \frac{H}{\dot
\sigma} \delta s \right] \;, \label{zetadefgen}
 \eeq
where $\delta \sigma$ is evaluated on a uniform flat hypersurface.
The evolution of $\zeta$ is sourced by first and second order
perturbations of $\delta s$. On super-Hubble scales it
reads\footnote{This equation corresponds to Eq.~(221) of
  \cite{Langlois:2006vv}. Note however that in \texttt{v1} and
  \texttt{v2}
of the \texttt{arXiv}
  and in the published version on JCAP,
the last term inside the
  bracket is missing in this equation. We thank S\'ebastien Renaux-Petel and Gianmassimo Tasinato for pointing out this
  omission.}
 \beq
 \label{zetadot}
 \dot \zeta = - \frac{H}{\dot \sigma^2} \left[ 2 \dot \theta \dot\sigma
  \delta s - (V_{,ss} + 4 \dot \theta^2) \delta s^2 +
  \frac{V_{,\sigma}}{\dot \sigma}\delta s \dot {\delta s} \right] \;,
 \eeq
where $\dot \theta = - V_{,s}/\dot \sigma$, with
$V_{,s}=-\sin\theta V_{,\phi}+\cos \theta V_{,\chi}$ and $V_{,ss}
= V_{,\phi \phi} \sin^2 \theta - 2V_{,\phi \chi} \cos \theta \sin
\theta + V_{,\chi \chi} \cos^2 \theta$. The entropy field
perturbation  $\delta s$ evolves independently on super-Hubble
scales and its evolution equation reads,
 \beq
\ddot{\delta s} + 3 H \dot{\delta s} + (V_{,ss} + 3 \dot \theta^2)
\delta s = - \frac{\dot \theta}{\dot \sigma} \delta s^2 -
\frac{2}{\dot \sigma} \left(\ddot \theta + \dot \theta
  \frac{V_{,\sigma}}{\dot \sigma} -\frac{3}{2} H \dot \theta  \right)
\delta s \dot{\delta s} - \left(\frac{1}{2} V_{,sss} - 5 \frac{\dot
    \theta}{\dot \sigma} V_{,ss} - 9 \frac{\dot \theta^3}{\dot \sigma}
\right) \delta s^2 \;,
 \eeq
where $V_{,\sigma} = \cos \theta V_{,\phi} + \sin \theta V_{, \chi}$
and $V_{,s ss} = - V_{,\phi \phi \phi} \sin^3 \theta + 3V_{,\phi
\phi \chi} \cos \theta \sin^2 \theta - 3V_{,\phi \chi \chi} \cos^2
\theta \sin \theta  + V_{,\chi \chi \chi} \cos^3 \theta$.

Given these evolution equations we expect that their solutions can
be written as
 \beq
\zeta = \zeta_* + T_{\zeta }^{(1)} \delta s_* + T_{\zeta }^{(2)}
\delta s_*^2\;, \label{gensolzeta}
 \eeq
 \beq
  \delta s = T^{(1)}_{\delta s} \delta s_* +
T^{(2)}_{\delta s} \delta s_*^2 \;, \label{gensols}
 \eeq
 where $T_{\zeta,\delta s}^{(1)}$ and $T_{\zeta,\delta s}^{(2)}$ are the
first and second order transfer functions, for $\zeta$ and $\delta
s$, respectively, and $\zeta_*$, $\delta s_*$ are their initial
conditions at Hubble exit. Equation~(\ref{zetarad2fieldsalpha}) is
an example of the general solution (\ref{gensolzeta}), while
Eq.~(\ref{Scdm2fieldsalpha}) is an example of (\ref{gensols})
rewritten in terms of the CDM entropy perturbation $\S_{\rm m}$.
%

Let $\epsilon$ be the standard slow-roll parameter,
and let us define the mass slow-roll
parameters $\eta_{ij}=V_{,ij}/(3H^2)$ and  $\eta_{\sigma s} =
(\eta_{\chi \chi} - \eta_{\phi \phi}) \cos \theta \sin \theta +
\eta_{\phi \chi} (\cos^2 \theta - \sin^2 \theta)$, $ \eta_{s s} =
\eta_{\phi \phi} \sin^2 \theta - 2\eta_{\phi \chi} \cos \theta
\sin \theta + \eta_{\chi \chi} \cos^2 \theta $. By using $\dot
\theta = - H \eta_{\sigma s}$,
which follows from the time derivative of
$\tan\theta=\dot\chi/\dot\phi\simeq V_{,\chi}/V_{,\phi}$,
 and $ \dot {\delta s} = - H
\eta_{ss} \delta s$, one can rewrite the second order definitions
of $\delta \sigma$ and $\delta s$, Eq.~(\ref{defrotsec}), and the
definition of $\zeta$, Eq.~(\ref{zetadefgen}), in terms of the
slow-roll parameters,
 \beq
 \delta \sigma = \delta \sigma^{(1)} - \frac{1}{2
\sqrt{2 \epsilon} \mP}\; \eta_{ss}\delta s^2 \;, \qquad
 \delta s = \delta s^{(1)} + \frac{1}{ \sqrt{2 \epsilon}\mP} \left(
\eta_{ss}{\delta s} +\frac{\eta_{\sigma s}}{2} \delta \sigma
\right)\delta \sigma \;, \label{deftotal}
\eeq
and
 \beq
 \label{zetastar-general}
 \zeta = -\frac{1}{\sqrt{2
\epsilon}\mP} \left\{ \delta \sigma- \frac{1}{\sqrt{2\epsilon}\mP}
\left[ \left(\epsilon-\frac{\eta_{\sigma \sigma}}{2} \right)\delta
\sigma^2- {\eta_{\sigma s}} \delta \sigma \delta s\right] \right\}
\;.
 \eeq
Note that in these three definitions, the non-linear terms on the
right hand side are
slow-roll suppressed with respect to the linear terms.
If evaluated at
Hubble crossing, when the slow-roll parameters are small, they
lead to contributions to the intrinsic non-Gaussianities of
$\delta \sigma_*$, $\delta s_*$, and $\zeta_*$ which are small.

For completeness, we compute here the 3-point correlation
functions of $\delta \sigma_*$ and $\delta s_*$.
From the results of
\cite{Seery:2005gb}, namely
\beq
\langle \delta \varphi_*^I(\vec k_1) \delta\varphi_*^{J}(\vec
k_2)\delta\varphi_*^{K}(\vec k_3)\rangle =(2 \pi)^3 \delta
(\Sigma_i \vec{k}_i) \frac{ H^4_*}{16 \mP^2}\sum_{\rm perms}\frac{\dot\varphi_*^I\delta^{JK}}{H_* \Pi_ik_i^3}{\cal M}(k_1,k_2,k_3)\;,
\eeq
with
\beq
{\cal M}(k_1,k_2,k_3)\equiv -k_1 k_2^2-4\frac{k_2^2k_3^2}{k_t}+\frac{1}{2}k_1^3+\frac{k_2^2k_3^2}{k_t^2}(k_2-k_3),
\quad k_t\equiv k_1+k_2+k_3\;,
\eeq
one  can compute the 3-point correlators of
$\delta \sigma^{(1)}$ and $\delta s^{(1)}$, simply by using the change of basis in field space (\ref{defrotlin}).
 We find that $\langle
\delta \sigma_*^{(1)} \delta \sigma_*^{(1)}\delta
\sigma_*^{(1)}\rangle$ is the same as for a single scalar field
\cite{Maldacena:2002vr}, whereas $\langle \delta \sigma_*^{(1)} \delta
\sigma_*^{(1)}\delta s_*^{(1)}\rangle$ and $\langle \delta
s_*^{(1)} \delta s_*^{(1)}\delta s_*^{(1)}\rangle$ vanish, simply because $\dot s=0$. We also find
\beq
\langle \delta s_*^{(1)}(\vec k_1) \delta s_*^{(1)}(\vec
k_2)\delta \sigma_*^{(1)}(\vec k_3)\rangle =(2 \pi)^3 \delta
(\Sigma_i \k_i) \frac{\sqrt{\epsilon_*} H^4_*}{8\sqrt{2} \mP}
 \left( \frac{k_3^3 -k_3(k_1^2+k_2^2)}{\Pi_i k_i^3}
- 8 \frac{k_1^2 k_2^2}{k_t \Pi_i k_i^3} \right)\;,
\label{sssigma}
\eeq
and similar expressions for $\langle \delta s_*^{(1)}(\vec k_1) \delta \sigma_*^{(1)}(\vec
k_2)\delta s_*^{(1)}(\vec k_3)\rangle$ and $\langle \delta \sigma_*^{(1)}(\vec k_1) \delta s_*^{(1)}(\vec
k_2)\delta s_*^{(1)}(\vec k_3)\rangle$ by relabeling appropriately the $k_i$ appearing on the right hand side of
(\ref{sssigma}).

Taking also into account  the second order parts of the adiabatic and entropy perturbations given in Eq.~(\ref{deftotal}),
 we eventually find
for the full 3-point functions
\bea
\langle \delta \sigma_*(\vec k_1) \delta \sigma_*(\vec k_2)\delta
\sigma_*(\vec k_3)\rangle &=& (2 \pi)^3 \delta (\Sigma_i \vec{k}_i)
\frac{\sqrt{\epsilon_*} H^4_*}{8\sqrt{2} \mP}
 \left( \frac{\Sigma_i k_i^3 - \Sigma_{i \neq j} k_i
k_j^2 }{\Pi_ik_i^3} -8 \frac{\Sigma_{i>j} k_i^2
k_j^2}{k_t \Pi_ik_i^3}  \right)\;, \\
\langle \delta \sigma_*(\vec k_1) \delta \sigma_*(\vec k_2)\delta
s_*(\vec k_3)\rangle &=& (2 \pi)^3 \delta (\Sigma_i \vec{k}_i)
\frac{\eta_{\sigma s*} H^4_*}{8\sqrt{2{\epsilon_*}} \mP} \frac{1}{k_1^3 k_2^3} \;,\\
\langle \delta s_*(\vec k_1) \delta s_*(\vec k_2)\delta
\sigma_*(\vec k_3)\rangle &=&
(2 \pi)^3 \delta (\Sigma_i \vec k_i) \frac{
H^4_*}{8\sqrt{2{\epsilon_*}} \mP}
\left[\epsilon_*
\left( \frac{k_3^3 -k_3(k_1^2+k_2^2)}{\Pi_i k_i^3}
- 8 \frac{k_1^2 k_2^2}{k_t \Pi_i k_i^3} \right)
\right.
\cr
&&\left. +\eta_{ss*}
 \left( \frac{-k_3^3 +2
k_1^3+2k_2^3}{\Pi_i k_i^3}
 \right)\right]\;, \\
\langle \delta s_*(\vec k_1) \delta s_*(\vec k_2)\delta s_*(\vec
k_3)\rangle &=& 0\;.
\eea
This shows that the intrinsic non-Gaussianities of $\delta
\sigma_*$, $\delta s_*$ and, as a consequence of
Eq.~(\ref{zetastar-general}), of $\zeta_*$ are all small
for slow-roll models.

\section{Primordial power spectra and bispectra}
\label{app:primordial}

At first-order the expressions in Eqs.~(\ref{zetarfinal})
and~(\ref{Smfinal}) for the primordial curvature and isocurvature
perturbations in terms of the field perturbations during inflation
give the power spectra of the primordial perturbations at leading
order
\begin{eqnarray}
 \langle \zetar(\k_1) \zetar(\k_2) \rangle &=& (2\pi)^3
 \delta^3(\k_1+\k_2) \left\{ N_{,\sigma}^2
 P_\sigma(k_1) + 2 N_{,\sigma} N_{,s} C_{s\sigma}(k_1) + N_{,s}^2 P_s(k_1) \right\}  \,,
 \\
\frac13 \langle \zetar(\k_1) \Sm(\k_2) \rangle &=& (2\pi)^3
 \delta^3(\k_1+\k_2) \left\{ N_{,s}
\Delta N_{,s}
 P_s(k_1) + N_{,\sigma} \Delta N_{,s} C_{s\sigma}(k_1) \right\}  \,,
 \\
\frac19 \langle \Sm(\k_1) \Sm(\k_2) \rangle &=& (2\pi)^3
 \delta^3(\k_1+\k_2) \Delta N_{,s}^2
P_s(k_1)  \,,
\end{eqnarray}
where at horizon-crossing during inflation we have
\begin{eqnarray}
 \langle \delta \sigma_*(\k_1) \delta \sigma_*(\k_2) \rangle
 &=& (2\pi)^3 \delta(\k_1+\k_2) P_\sigma(k_1)  \,,\\
\langle \delta s_*(\k_1) \delta s_*(\k_2) \rangle
 &=& (2\pi)^3 \delta(\k_1+\k_2) P_s(k_1)  \,,\\
\langle \delta \sigma_*(\k_1) \delta s_*(\k_2) \rangle
 &=& (2\pi)^3 \delta(\k_1+\k_2) C_{s\sigma}(k_1)  \,.
\end{eqnarray}

At second-order Eqs.~(\ref{zetarfinal}) and~(\ref{Smfinal}) give the
leading order bispectra for the primordial curvature and
isocurvature perturbations and their correlations,
\bea
\langle \zeta_{\rm r}(\k_1)\zeta_{\rm r}(\k_2)\zeta_{\rm r}(\k_3) \rangle
&=& N_{,I}N_{,J}N_{,K} \langle \delta \varphi_*^I(\k_1)
\delta \varphi_*^J(\k_2) \delta \varphi_*^K(\k_3) \rangle \nonumber \\
&+&  (2\pi)^3 \delta(\Sigma_i \k_i)
N_{,I} N_{,J} N_{,KL} [C^{IK}(k_1)
C^{JL}(k_2) + 2 \ {\rm perms}]\;, \\
\frac13
\langle \zeta_{\rm r}(\k_1)\zeta_{\rm r}(\k_2){\cal S}_{\rm m}(\k_3) \rangle
&=& N_{,I}N_{,J}\Delta N_{,K} \langle \delta \varphi_*^I(\k_1)
\delta \varphi_*^J(\k_2) \delta \varphi_*^K(\k_3) \rangle \nonumber \\
&+ &  (2\pi)^3 \delta(\Sigma_i \k_i)
\Big\{ N_{,I} N_{,J} \Delta N_{,KL}
 C^{IK}(k_1) C^{JL}(k_2)  \nonumber \\
&+& N_{,IJ} N_{,K} \Delta N_{,L} [C^{JK}(k_1) + C^{JK}(k_2)
]C^{IL}(k_3) \Big\}\;, \\
\frac19
\langle {\cal S}_{\rm m}(\k_1){\cal S}_{\rm m}(\k_2)\zeta_{\rm r}(\k_3) \rangle
&=& \Delta N_{,I} \Delta N_{,J} N_{,K} \langle \delta \varphi_*^I(\k_1)
\delta \varphi_*^J(\k_2) \delta \varphi_*^K(\k_3) \rangle \nonumber \\
&+ &  (2\pi)^3 \delta(\Sigma_i \k_i)  \Big\{ \Delta N_{,I} \Delta
N_{,J} N_{,KL}
 C^{IK}(k_1) C^{JL}(k_2)  \nonumber \\
&+&  \Delta N_{,IJ} \Delta N_{,K} N_{,L}  [C^{JK}(k_1) + C^{JK}(k_2)
]C^{IL}(k_3) \Big\} \;, \\
\frac{1}{27}
\langle {\cal S}_{\rm m}(\k_1){\cal S}_{\rm m}(\k_2) {\cal S}_{\rm
  m}(\k_3)  \rangle
&=& \Delta N_{,I}\Delta N_{,J}\Delta N_{,K} \langle \delta \varphi_*^I(\k_1)
\delta \varphi_*^J(\k_2) \delta \varphi_*^K(\k_3) \rangle \nonumber \\
&+&  (2\pi)^3 \delta(\Sigma_i \k_i)
\Delta N_{,I} \Delta N_{,J} \Delta
N_{,KL} [C^{IK}(k_1)
C^{JL}(k_2) + 2 \ {\rm perms}]\;,
\eea
where $C^{II} (k) \equiv P_I(k)$.

At leading order in a slow-roll expansion the adiabatic and entropy
field perturbations are independent Gaussian random fields, with
\cite{Byrnes:2006fr,Lalak:2007vi}
 \begin{equation}
  P_s(k) \simeq P_\sigma(k) \simeq P_*(k) = \frac{H_*}{2k^3} \,,
  \qquad
  C_{s\sigma}(k) \simeq 0 \,,
 \end{equation}
and the primordial bispectra simplify considerably in the slow-roll limit,
where we drop the terms in brackets in Eqs.~(\ref{zetarfinal}) and
(\ref{Smfinal}), to give
\begin{eqnarray}
 \langle \zetar(\k_1) \zetar(\k_2) \zetar(\k_3) \rangle
 &\simeq&
 (2\pi)^3 \delta(\Sigma_i \k_i) N_{,s}^2 N_{ss} \left[ P_*(k_1) P_*(k_2) + 2 \ {\rm perms}\right]
\,, \\
\frac13 \langle \zetar(\k_1) \zetar(\k_2) \Sm(\k_3) \rangle
 &\simeq& (2\pi)^3 \delta(\Sigma_i \k_i) \Big\{
  N_{,s} N_{,ss} \Delta N_{,s}
  \left[ P_*(k_1)+ P_*(k_2)  \right] P_*(k_3)
   \nonumber\\
  &&   + N_{,s}^2 \Delta N_{,ss} P_*(k_1) P_*(k_2)
  \Big\}
\,, \\
\frac19 \langle \Sm(\k_1) \Sm(\k_2) \zeta_{\rm r}(\k_3)  \rangle
 &\simeq& (2\pi)^3 \delta(\Sigma_i \k_i) \Big\{
  N_{,s} \Delta N_{,s} \Delta N_{,ss}
  \left[ P_*(k_1) + P_*(k_2)  \right]P_*(k_3)
 \nonumber\\
  && + N_{,ss} \Delta N_{,s}^2 P_*(k_1) P_*(k_2)  \Big\}  \,,
\\
\frac{1}{27} \langle \Sm(\k_1) \Sm(\k_2) \Sm(\k_3) \rangle
 &\simeq&  (2\pi)^3 \delta(\Sigma_i \k_i)
\Delta N_{,s}^2 \Delta N_{,ss} \left[ P_*(k_1) P_*(k_2)  + 2 \ {\rm perms}\right] \,.
\end{eqnarray}


\footnotesize
\begin{thebibliography}{99}

\bibitem{Linde:1996gt}
  A.~D.~Linde and V.~F.~Mukhanov,
  ``Nongaussian isocurvature perturbations from inflation,''
  Phys.\ Rev.\  D {\bf 56}, 535 (1997)
  [arXiv:astro-ph/9610219].

\bibitem{Boubekeur:2005fj}
  L.~Boubekeur and D.~H.~Lyth,
  Phys.\ Rev.\  D {\bf 73}, 021301 (2006)
  [arXiv:astro-ph/0504046].


\bibitem{deltaN}
  M.~Sasaki and E.~D.~Stewart,
  ``A General analytic formula for the spectral index of the density
  perturbations produced during inflation,''
  Prog.\ Theor.\ Phys.\  {\bf 95}, 71 (1996)
  [arXiv:astro-ph/9507001];
  A.~A.~Starobinsky,
 ``Multicomponent de Sitter (inflationary) stages and the generation
 of perturbations,''
  JETP Lett.\  {\bf 42}, 152 (1985)
  [Pisma Zh.\ Eksp.\ Teor.\ Fiz.\  {\bf 42}, 124 (1985)].

\bibitem{Lyth:2005fi}
  D.~H.~Lyth and Y.~Rodriguez,
  ``The inflationary prediction for primordial non-gaussianity,''
  Phys.\ Rev.\ Lett.\  {\bf 95}, 121302 (2005)
  [arXiv:astro-ph/0504045].


\bibitem{curvaton}
  K.~Enqvist and M.~S.~Sloth,
  ``Adiabatic CMB perturbations in pre big bang string cosmology,''
  Nucl.\ Phys.\  B {\bf 626}, 395 (2002)
  [arXiv:hep-ph/0109214];
  D.~H.~Lyth and D.~Wands,
  ``Generating the curvature perturbation without an inflaton,''
  Phys.\ Lett.\  B {\bf 524}, 5 (2002)
  [arXiv:hep-ph/0110002];
  T.~Moroi and T.~Takahashi,
  ``Effects of cosmological moduli fields on cosmic microwave background,''
  Phys.\ Lett.\  B {\bf 522}, 215 (2001)
  [Erratum-ibid.\  B {\bf 539}, 303 (2002)]
  [arXiv:hep-ph/0110096].

\bibitem{Lyth:2002my}
  D.~H.~Lyth, C.~Ungarelli and D.~Wands,
  ``The primordial density perturbation in the curvaton scenario,''
  Phys.\ Rev.\  D {\bf 67}, 023503 (2003)
  [arXiv:astro-ph/0208055].

\bibitem{Gordon:2002gv}
  C.~Gordon and A.~Lewis,
  ``Observational constraints on the curvaton model of inflation,''
  Phys.\ Rev.\  D {\bf 67}, 123513 (2003)
  [arXiv:astro-ph/0212248].

\bibitem{beltran}
  M. Beltran,
  ``Isocurvature, non-gaussianity and the curvaton model,''
  Phys.\ Rev.\  D {\bf 78}, 023530 (2008)
  [arXiv:0804.1097 [astro-ph]].

\bibitem{Langlois:2004nn}
  D.~Langlois and F.~Vernizzi,
  ``Mixed inflaton and curvaton perturbations,''
  Phys.\ Rev.\  D {\bf 70}, 063522 (2004)
  [arXiv:astro-ph/0403258].

\bibitem{Ferrer:2004nv}
  F.~Ferrer, S.~Rasanen and J.~Valiviita,
  ``Correlated isocurvature perturbations from mixed inflaton-curvaton
  decay,''
  JCAP {\bf 0410}, 010 (2004)
  [arXiv:astro-ph/0407300].

\bibitem{Polarski:1992dq}
  D.~Polarski and A.~A.~Starobinsky,
  ``Spectra of perturbations produced by double inflation with an intermediate
  matter dominated stage,''
  Nucl.\ Phys.\ B {\bf 385}, 623 (1992).

\bibitem{Gordon:2000hv}
  C.~Gordon, D.~Wands, B.~A.~Bassett and R.~Maartens,
  ``Adiabatic and entropy perturbations from inflation,''
  Phys.\ Rev.\  D {\bf 63}, 023506 (2001)
  [arXiv:astro-ph/0009131].

\bibitem{GrootNibbelink:2001qt}
  S.~Groot Nibbelink and B.~J.~W.~van Tent,
  ``Scalar perturbations during multiple field slow-roll inflation,''
  Class.\ Quant.\ Grav.\  {\bf 19}, 613 (2002)
  [arXiv:hep-ph/0107272].

\bibitem{NLdec}
  G.~I.~Rigopoulos, E.~P.~S.~Shellard and B.~W.~van Tent,
  ``Non-linear perturbations in multiple-field inflation,''
  Phys.\ Rev.\ D {\bf 73}, 083521 (2006)
  [arXiv:astro-ph/0504508].


\bibitem{Langlois:2006vv}
  D.~Langlois and F.~Vernizzi,
  ``Nonlinear perturbations of cosmological scalar fields,''
  JCAP {\bf 0702}, 017 (2007)
  [arXiv:astro-ph/0610064].

\bibitem{DBI}
 D.~Langlois, S.~Renaux-Petel, D.~A.~Steer and T.~Tanaka,
  ``Primordial fluctuations and non-Gaussianities in multi-field DBI
  inflation,'', Phys.\ Rev.\ Lett.\  {\bf 101},  061301 (2008)
  [arXiv:0804.3139 [hep-th]];
  D.~Langlois, S.~Renaux-Petel, D.~A.~Steer and T.~Tanaka,
  ``Primordial perturbations and non-Gaussianities in DBI and general
  multi-field inflation,''  Phys.\ Rev.\ D  {\bf 78}, 063523 (2008)
  [arXiv:0806.0336 [hep-th]].

\bibitem{Kawasaki:2008sn}
  M.~Kawasaki, K.~Nakayama, T.~Sekiguchi, T.~Suyama and F.~Takahashi,
  ``Non-Gaussianity from isocurvature perturbations,''
  arXiv:0808.0009 [astro-ph].

\bibitem{Kawasaki:2008jy}
  M.~Kawasaki, K.~Nakayama and F.~Takahashi,
  ``Non-Gaussianity from Baryon Asymmetry,''
  arXiv:0809.2242 [hep-ph].



\bibitem{Langlois:2005ii}
  D.~Langlois and F.~Vernizzi,
  ``Evolution of nonlinear cosmological perturbations,''
  Phys.\ Rev.\ Lett.\  {\bf 95}, 091303 (2005)
  [arXiv:astro-ph/0503416].

\bibitem{Langlois:2005qp}
  D.~Langlois and F.~Vernizzi,
  ``Conserved nonlinear quantities in cosmology,''
  Phys.\ Rev.\ D {\bf 72}, 103501 (2005)
  [arXiv:astro-ph/0509078].

\bibitem{Langlois:2006iq}
  D.~Langlois and F.~Vernizzi,
  ``Nonlinear perturbations for dissipative and interacting relativistic
  fluids,''
  JCAP {\bf 0602}, 014 (2006)
  [arXiv:astro-ph/0601271].

\bibitem{Lyth:2004gb}
  D.~H.~Lyth, K.~A.~Malik and M.~Sasaki,
  ``A general proof of the conservation of the curvature perturbation,''
  JCAP {\bf 0505}, 004 (2005)
  [arXiv:astro-ph/0411220].

\bibitem{Sasaki:2006kq}
  M.~Sasaki, J.~Valiviita and D.~Wands,
  ``Non-gaussianity of the primordial perturbation in the curvaton model,''
  Phys.\ Rev.\  D {\bf 74}, 103003 (2006)
  [arXiv:astro-ph/0607627].

\bibitem{Polarski:1994rz}
  D.~Polarski and A.~A.~Starobinsky,
  ``Isocurvature perturbations in multiple inflationary models,''
  Phys.\ Rev.\  D {\bf 50}, 6123 (1994)
  [arXiv:astro-ph/9404061].

\bibitem{RC}
  F.~Vernizzi,
  ``On the conservation of second-order cosmological perturbations in a  scalar
  field dominated Universe,''
  Phys.\ Rev.\ D {\bf 71}, 061301 (2005)
  [arXiv:astro-ph/0411463].

\bibitem{Seery:2005gb}
  D.~Seery and J.~E.~Lidsey,
  ``Primordial non-gaussianities from multiple-field inflation,''
  JCAP {\bf 0509}, 011 (2005)
  [arXiv:astro-ph/0506056].

\bibitem{Malik:2002jb}
  K.~A.~Malik, D.~Wands and C.~Ungarelli,
  ``Large-scale curvature and entropy perturbations for multiple interacting
  fluids,''
  Phys.\ Rev.\  D {\bf 67}, 063516 (2003)
  [arXiv:astro-ph/0211602].

\bibitem{Malik:2006pm}
  K.~A.~Malik and D.~H.~Lyth,
  ``A numerical study of non-gaussianity in the curvaton scenario,''
  JCAP {\bf 0609}, 008 (2006)
  [arXiv:astro-ph/0604387].

\bibitem{Vernizzi:2005fx}
  F.~Vernizzi,
  ``Generating cosmological perturbations with mass variations,''
  Nucl.\ Phys.\ Proc.\ Suppl.\  {\bf 148}, 120 (2005)
  [arXiv:astro-ph/0503175].

\bibitem{Maldacena:2002vr}
  J.~Maldacena,
  ``Non-Gaussian features of primordial fluctuations in single field
  inflationary models,''
  JHEP {\bf 0305}, 013 (2003)
  [arXiv:astro-ph/0210603].

\bibitem{Acquaviva:2002ud}
  V.~Acquaviva, N.~Bartolo, S.~Matarrese and A.~Riotto,
  ``Second-order cosmological perturbations from inflation,''
  Nucl.\ Phys.\  B {\bf 667}, 119 (2003)
  [arXiv:astro-ph/0209156].

\bibitem{Ichikawa:2008iq}
  K.~Ichikawa, T.~Suyama, T.~Takahashi and M.~Yamaguchi,
  ``Non-Gaussianity, Spectral Index and Tensor Modes in Mixed Inflaton and Curvaton Models,''
  Phys.\ Rev.\  D {\bf 78}, 023513 (2008)
  [arXiv:0802.4138 [astro-ph]].


\bibitem{Bartolo:2003bz}
  N.~Bartolo, S.~Matarrese and A.~Riotto,
  ``Evolution of second-order cosmological perturbations and
  non-Gaussianity,''
  JCAP {\bf 0401}, 003 (2004)
  [arXiv:astro-ph/0309692].


\bibitem{Komatsu:2001rj}
  E.~Komatsu and D.~N.~Spergel,
  ``Acoustic signatures in the primary microwave background bispectrum,''
  Phys.\ Rev.\  D {\bf 63}, 063002 (2001)
  [arXiv:astro-ph/0005036].

\bibitem{Gupta:2003jc}
  S.~Gupta, K.~A.~Malik and D.~Wands,
  Phys.\ Rev.\  D {\bf 69}, 063513 (2004)
  [arXiv:astro-ph/0311562].

\bibitem{Weinberg:2004kf}
  S.~Weinberg,
  ``Must cosmological perturbations remain non-adiabatic after multi-field
  inflation?,''
  Phys.\ Rev.\  D {\bf 70}, 083522 (2004)
  [arXiv:astro-ph/0405397].



\bibitem{Komatsu:2008hk}
  E.~Komatsu {\it et al.}  [WMAP Collaboration],
  ``Five-Year Wilkinson Microwave Anisotropy Probe (WMAP)
  Observations:Cosmological Interpretation,''
  arXiv:0803.0547 [astro-ph].

\bibitem{Garcia-Bellido:1995kc}
  J.~Garcia-Bellido and D.~Wands,
  ``General relativity as an attractor in scalar - tensor stochastic
  inflation,''
  Phys.\ Rev.\ D {\bf 52}, 5636 (1995)
  [arXiv:gr-qc/9503049].

\bibitem{Mukhanov:1997fw}
  V.~F.~Mukhanov and P.~J.~Steinhardt,
  ``Density perturbations in multifield inflationary models,''
  Phys.\ Lett.\  B {\bf 422}, 52 (1998)
  [arXiv:astro-ph/9710038].


\bibitem{Vernizzi:2006ve}
  F.~Vernizzi and D.~Wands,
  ``Non-Gaussianities in two-field inflation,''
  JCAP {\bf 0605}, 019 (2006)
  [arXiv:astro-ph/0603799].



\bibitem{Alabidi:2005qi}
  L.~Alabidi and D.~H.~Lyth,
  ``Inflation models and observation,''
  JCAP {\bf 0605}, 016 (2006)
  [arXiv:astro-ph/0510441].

\bibitem{Rigopoulos:2005ae}
  G.~I.~Rigopoulos, E.~P.~S.~Shellard and B.~J.~W.~van Tent,
  ``Large non-Gaussianity in multiple-field inflation,''
  Phys.\ Rev.\  D {\bf 73}, 083522 (2006)
  [arXiv:astro-ph/0506704].

\bibitem{Bernardeau:2002jy}
  F.~Bernardeau and J.~P.~Uzan,
  ``Non-Gaussianity in multi-field inflation,''
  Phys.\ Rev.\  D {\bf 66}, 103506 (2002)
  [arXiv:hep-ph/0207295].


\bibitem{Langlois:1999dw}
  D.~Langlois,
  ``Correlated adiabatic and isocurvature perturbations from double
  inflation,''
  Phys.\ Rev.\ D {\bf 59}, 123512 (1999)
  [arXiv:astro-ph/9906080].

\bibitem{Wands:2000dp}
  D.~Wands, K.~A.~Malik, D.~H.~Lyth and A.~R.~Liddle,
  ``A new approach to the evolution of cosmological perturbations on large
  scales,''
  Phys.\ Rev.\  D {\bf 62}, 043527 (2000)
  [arXiv:astro-ph/0003278].

\bibitem{Bartolo:2004if}
  N.~Bartolo, E.~Komatsu, S.~Matarrese and A.~Riotto,
  ``Non-Gaussianity from inflation: Theory and observations,''
  Phys.\ Rept.\  {\bf 402}, 103 (2004)
  [arXiv:astro-ph/0406398].

\bibitem{Byrnes:2006fr}
  C.~T.~Byrnes and D.~Wands,
  ``Curvature and isocurvature perturbations from two-field inflation in a
  slow-roll expansion,''
  Phys.\ Rev.\  D {\bf 74}, 043529 (2006)
  [arXiv:astro-ph/0605679].

\bibitem{Lalak:2007vi}
  Z.~Lalak, D.~Langlois, S.~Pokorski and K.~Turzynski,
  ``Curvature and isocurvature perturbations in two-field inflation,''
  JCAP {\bf 0707}, 014 (2007)
  [arXiv:0704.0212 [hep-th]].

\end{thebibliography}
\end{document}